\documentclass{cmspaper}
\usepackage{epsfig}
\usepackage{amssymb}
\begin{document}
\begin{titlepage}
\cmsnote{2001/042}
\date{\today}
\title{Observability of MSSM Higgs bosons via sparticle
decay modes in CMS}
\begin{Authlist}
Filip~Moortgat\Aref{a}
\Instfoot{antw}{University of Antwerpen, Wilrijk, Belgium}
Salavat~Abdullin\Aref{b}
\Instfoot{itep}{University of Maryland, College Park, USA}
Daniel~Denegri
\Instfoot{sac}{DAPNIA Saclay, France - CERN, Geneva, Switzerland}
\end{Authlist}
\Anotfoot{a}{Research Assistant of the Fund for Scientific Research -
Flanders (Belgium)}
\Anotfoot{b}{On leave of absence from ITEP, Moscow, Russia}
\begin{abstract}
\noindent
We discuss the possibilities to observe the decays of heavy SUSY Higgs bosons
into supersymmetric particles at the LHC. Such an observation would be of interest either
in a discovery search if sparticle modes are the dominant ones, or in a 
study of additional decay modes, bringing information on the SUSY scenario
potentially at work. We will focus on the most promising channel 
where the heavy neutral Higgses decay into a pair of next-to-lightest neutralinos $\chi^0_2$,
followed by $\chi^0_2 \rightarrow l^+ l^- \chi^0_1$, thus leading to 
four isolated leptons + $E_T^{miss}$ as the main final state signature. A study with 
the CMS detector shows that the background (SM + SUSY) can be sufficiently 
suppressed and that in the mass region between $m_{A,H}$ $\sim$ 230 and 450 GeV, for low and
intermediate values of $\tan \beta$, the signal would be visible provided neutralinos 
and sleptons are light enough. 
\end{abstract}
\end{titlepage}
\setcounter{page}{2}
\section{Introduction}\label{chap:intro}
While the problem of electroweak symmetry breaking can be solved in the Standard Model (SM) by 
introducing one Higgs boson, the Minimal Supersymmetric Standard Model (MSSM) requires
five physical Higgses: a light CP-even ($h^0$), a heavy CP-even ($H^0$), a heavy CP-odd ($A^0$) and 
two charged Higgs bosons ($H^{\pm}$).
Therefore, the discovery of heavy neutral Higgs bosons would be a major breakthrough in 
verifying the supersymmetric nature of the fundamental theory, which is one of the main physics goals
of the Large Hadron Collider project.\\
The most promising channel to discover the heavy SUSY Higgses is the $A^0, H^0 \rightarrow \tau \tau$ \cite{ritva1}
channel, where both the leptonic and hadronic decays of the tau can be exploited.
This channel has been shown to cover large parts of the intermediate and high $\tan\beta$ region of the MSSM 
parameter space for an integrated luminosity of 30 $fb^{-1}$. 
For low values of $\tan\beta$, the coupling of the Higgs bosons to taus is not sufficiently enhanced 
and therefore this region is inaccessible for the $\tau\tau$ channel. 
\\
In all studies of the SM channels (meaning that the SUSY Higgses decay into
Standard Model particles), it is assumed that sparticles are too heavy to participate in the
decay process.
One should ask what would happen if some of the sparticles would be light and the decays 
of Higgs bosons into these SUSY particles would be kinematically allowed. 
Indeed, the existence of light neutralinos ($\chi^0$), charginos ($\chi^{\pm}$) and sleptons ($\tilde{l}$)
seems favoured by a large number of supersymmetric models in order to explain electroweak symmetry 
breaking without large fine-tuning \cite{kane}. Also recent experimental results (precision measurements 
at LEP2 \cite{alta}, muon $g - 2$ \cite{gmin}) may point towards the existence of light gauginos and sleptons.\\
Light SUSY particles may jeopardize the Higgs discovery potential of the 
SM channels, since their presence can drastically decrease the branching ratios of the Higgses into SM
particles. Furthermore, pair and cascade production of light sparticles becomes an extra background 
to the Higgs searches. 
On the other hand, Higgs bosons decaying into sparticles might open new possibilities to
explore regions of parameter space where SM decays would not be accessible \cite{baer1}. 
In this note we report on a study of this type of decay with the CMS detector. 
We will focus on the decay of the heavy neutral Higgses $H^0$ and $A^0$ into two
next-to-lightest neutralinos, 
with each of the neutralinos in turn decaying as $\chi^0_2 \rightarrow l^+ l^- \chi^0_1$, i.e.
into two (isolated) leptons + $E_T^{miss}$, 
so we get  
\begin{equation}
A^0, H^0 \rightarrow  \chi^0_2 \, \chi^0_2 \rightarrow 4 \, l^{\pm}  \; + \; X     \; \; \; \;
\, \, \, \, \, \, \, \, \, \, \, \, \,           (l\, =\, e,\, \mu)
\end{equation}
This results in a clear four lepton final state signature.
We will show that, as is often the case for supersymmetric channels, SUSY backgrounds are more 
difficult to suppress than the SM backgrounds.  
Of the latter, basically only $ZZ$ survives after requiring four isolated leptons.
Of the SUSY backgrounds, sneutrino pair production and sparticle cascade decay production of neutralinos 
are the most dangerous processes. 
Using a set of selection criteria as described in section 5, we can clearly distinguish the
signal from the background in the intermediate mass range 230 GeV $\lesssim$ $m_A$ $\lesssim$ 450 GeV and for low and
intermediate values of $\tan\beta$, depending on the values of the other MSSM parameters.  

The remainder of this note is organised as follows: first we study the behaviour of the 
relevant branching ratios. Then we describe the event generation, the signal versus background
discrimination methods, and the discovery potential of the channel in the $m_A$ - $\tan\beta$
plane. As a next step we investigate the effects of varying the other
MSSM parameter values. In the last section the results are summarized.

\section{Framework}\label{chap:frame}
The main difficulty in studying decay modes involving supersymmetric particles is the large 
amount of free parameters in the MSSM. Therefore most studies are carried out in the mSUGRA 
or GMSB context in order to reduce the number of free parameters; we will however stick to the 
more general MSSM framework, to avoid too many model dependent assumptions. As free
parameters, we take the mass of the CP-odd Higgs $m_A$, the Higgs VEV ratio
$\tan \beta$, the Higgsino mass parameter $\mu$, the bino mass parameter $M_1$, the wino
mass parameter $M_2$, the slepton mass $m_{\tilde{l}}$ and the squark/gluino mass $m_{\tilde{q},\tilde{g}}$. 
As a starting point for our studies, we will adopt the following framework:
\begin{2figures}{h}
  \resizebox{\linewidth}{1.1\linewidth}{\includegraphics{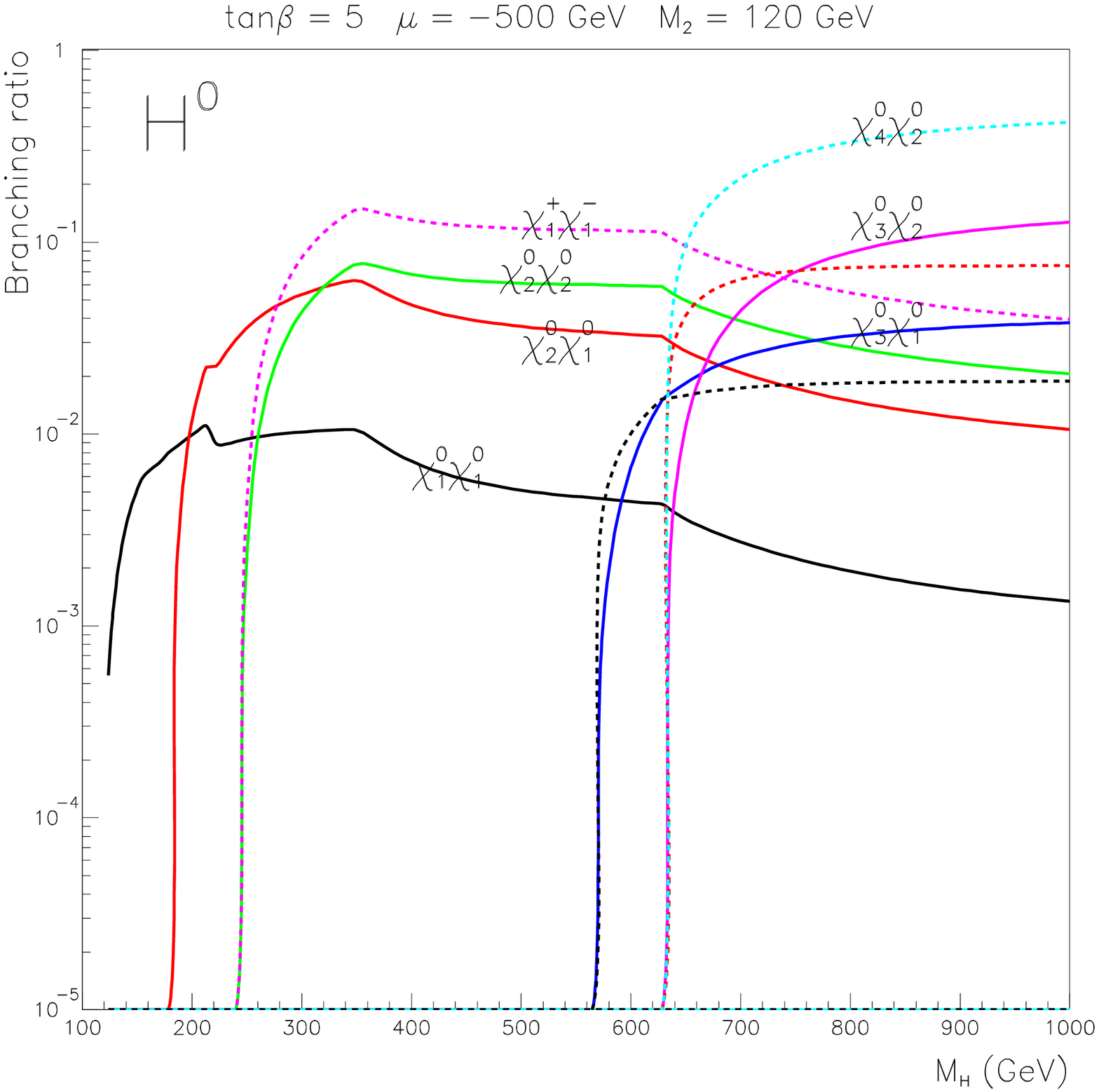}} &
  \resizebox{\linewidth}{1.1\linewidth}{\includegraphics{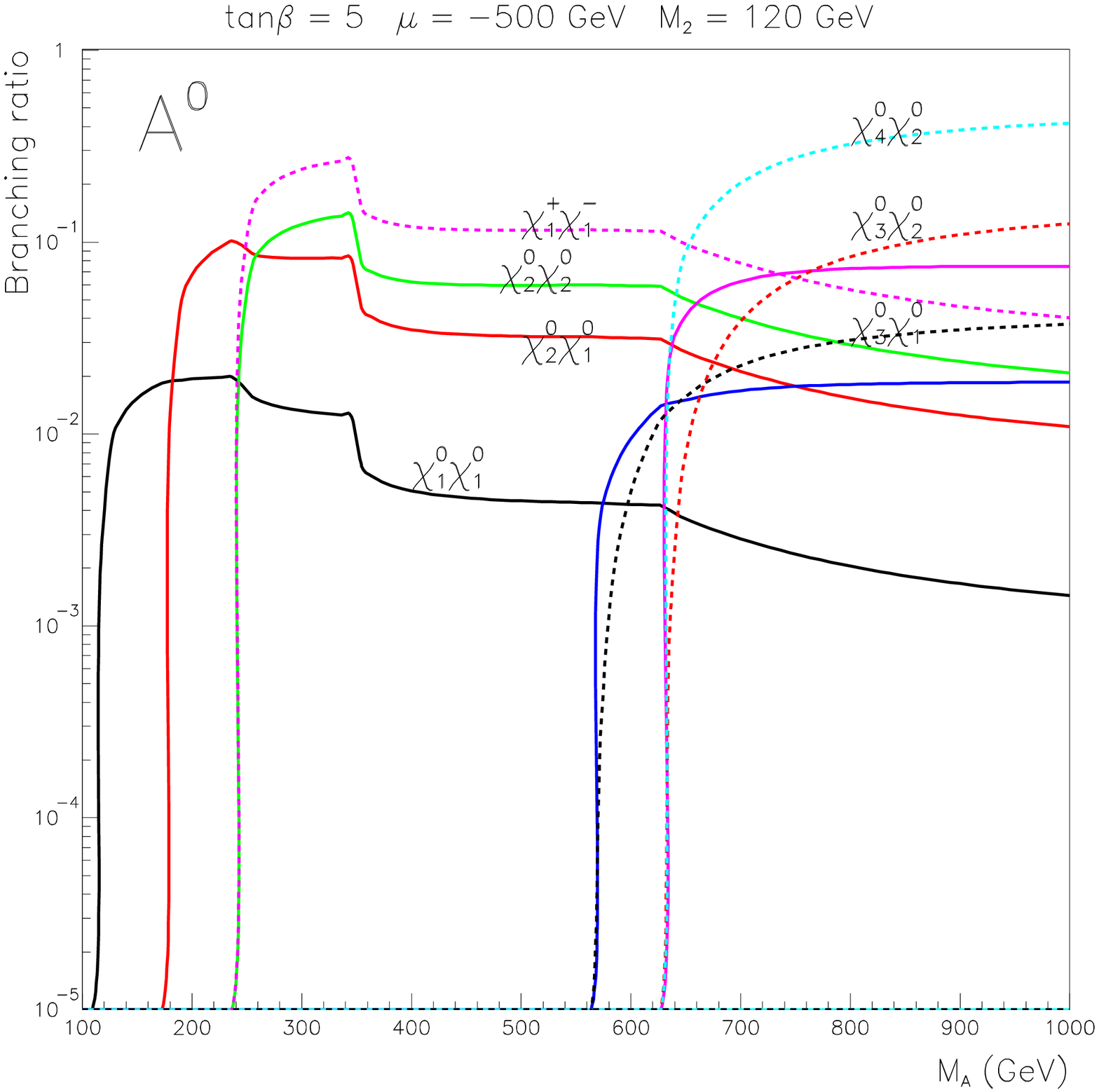}} \\
  \caption{Branching ratios of $H^0$ into neutralinos and charginos.}
  \label{fig:hd1} &
  \caption{Branching ratios of $A^0$ into neutralinos and charginos.} 
  \label{fig:hd2} \\
\end{2figures}
\begin{itemize}
\item we consider light neutralinos and charginos, above the LEP2 limits. 
Initially, we fix $M_1$ at 60 GeV, and using the renormalisation group relation $M_2$ $\approx$ 2 $M_1$, we can set $M_2$= 120 GeV. 
We take $M1<M2<|\mu|$. This large $\mu$ scenario is favoured in models where $\chi^0_1$ is the dark matter
candidate, like mSUGRA. In low $\mu$ scenarios, the decay of $\chi^0_2$ into leptons will be strongly
suppressed. For large values of $|\mu|$, $\chi^0_2$ is rather wino and $\chi^0_1$ is
bino-like. Therefore it approximately holds that $m_{\chi^0_1} \approx M_1$ and $m_{\chi^0_2} \approx M_2$.
The effects of varying these parameters will be discussed later on.
\item we also take sleptons to be light. In the most favourable case they would be lighter than $\chi^0_2$, thereby allowing
two-body decays into leptons. We will consider two scenarios: 
$m_{\tilde{l}}$ $<$ $m_{\chi^0_2}$, where real decays of neutralinos into sleptons are allowed 
and $m_{\tilde{l}}$ $>$ $m_{\chi^0_2}$, where only the virtual exchange is possible.
\item the masses of squarks and gluinos are kept at the 1 TeV scale. In the MSSM, it is natural 
that these sparticles are heavier than neutralinos and sleptons. In section 7, we will investigate
the effect of lowering the masses of squarks and gluinos.
\end{itemize}
These parameter values and domains for $\mu$, $M_1$, $M_2$,
$m_{\tilde{l}}$ and $m_{\tilde{q},\tilde{g}}$ will be used as default throughout this note. 
The exact values for $\mu$ and $m_{\tilde{l}}$ will be chosen after analysing and optimizing the $A^0, H^0 \rightarrow 4 \, l$ cross sections
through the MSSM parameter space. After establishing the visibility in this optimal point, we will scan the
area in $m_A - \tan\beta$ around it to see how far the discovery region reaches.
Effects of varying the initial SUSY parameter values will be discussed.

\section{Branching ratios in the MSSM parameter space}\label{chap:br}

In order to determine the regions in MSSM parameter space where sparticle decay modes may be 
accessible, we will first discuss the behaviour of the relevant branching ratios.
\subsection{Decay of the heavy Higgs bosons into neutralinos and charginos}
The package HDECAY \cite{hdecay} is used to study the supersymmetric decay modes $H^0,
A^0\rightarrow\chi\chi$ of the heavy Higgses. The $H^0$ and $A^0$ couple preferably to mixtures of gauginos and
higgsinos. The dominant MSSM parameters controlling this process are $m_A$, $tan\beta$, $M_2$ and $\mu$. \\
Figs. \ref{fig:hd1} and \ref{fig:hd2} show the decays of $H^0$ and $A^0$ 
into neutralinos and charginos, in the case where $M_1$ = 60 GeV$, M_2$~= 120 GeV, 
$\mu$ = -500 GeV and $\tan\beta$ = 5.
For $M_A$ $\lesssim$ 500 GeV, the probability for the heavy Higgses to decay into SUSY
particles can be as high as 20 \%.
In this mass region, the decay mode to $\chi_1^+\chi_1^-$ has the highest
branching ratio (BR), however it
produces a final state with only two leptons. This mode would be in competition with numerous SM and SUSY backgrounds.
The second best sparticle mode is $\chi^0_2 \chi^0_2$. This channel can provide four 
leptons and is thus a priori more appropriate for obtaining a good signal to
background ratio.
The $\chi^0_2 \chi^0_2$ threshold is determined by our choice of $M_2$ ($\approx
m_{\chi^0_2}$) = 120 GeV; the fall in BR at $M_A$ $>$ 350 GeV is caused by the opening of the $t\bar{t}$ mode. 
The partial decay widths into supersymmetric particles remain the same as for lower values of $M_A$, but due to the 
opening of the $t\bar{t}$ mode, the total decay width increases and 
the branching ratio into charginos/neutralinos decreases.
For values of $M_A$ $<$ 350 GeV, the BR of the CP-odd Higgs ($A^0$) into gauginos is 
substantially higher than in the CP-even ($H^0$) case. This is due to the fact
that for the CP-even Higgs more couplings to SM particles are allowed, thus leading 
to a larger total decay width and smaller BR's to sparticles. 
For high values of $m_A$ the BR's are about the same for $H^0$ and $A^0$ since one reaches the 
decoupling regime. Also for higher masses, other neutralino modes like $\chi^0_3 \chi^0_2$,
$\chi^0_4 \chi^0_2$ or $\chi^+_2 \chi^-_1$ may open up which will contribute to the four lepton signal.

\subsection{Decays of the next-to-lightest neutralino into leptons}

The next-to-lightest neutralino $\chi^0_2$ will decay into two fermions and the lightest 
neutralino: $\chi^0_2 \rightarrow f \bar{f} \chi^0_1$. These fermions will most often be quarks,
leading to two jets and missing $E_T$ in the final state. To obtain a clean signature, we will
only focus on the case where the  neutralino decays into two leptons
$\chi^0_2 \rightarrow l^+l^-\chi^0_1$, where $l$ = $e$ or $\mu$.
This process is determined by the bino, wino and higgsino mass parameters $M_1$, $M_2$, $\mu$, 
by $\tan\beta$ and by the slepton masses $m_{\tilde{l}}$.
If sleptons are heavier than the $\chi^0_2$, and as long as direct decays into a $Z^0$ boson are not allowed
(or suppressed), only three-body decays $\chi^0_2 \rightarrow l^+l^-\chi^0_1$ will contribute. 
These decays are mediated by virtual slepton and $Z^0$ exchange \cite{baer2}. Therefore it is more favourable to 
have light sleptons in order to have larger BR's. The decay branching ratios
can be rather sensitive to the MSSM parameters due to the fact that the $Z^0$ and slepton
exchange amplitudes may interfere constructively as well as destructively.
\begin{2figures}{ht}
  \resizebox{\linewidth}{!}{\includegraphics{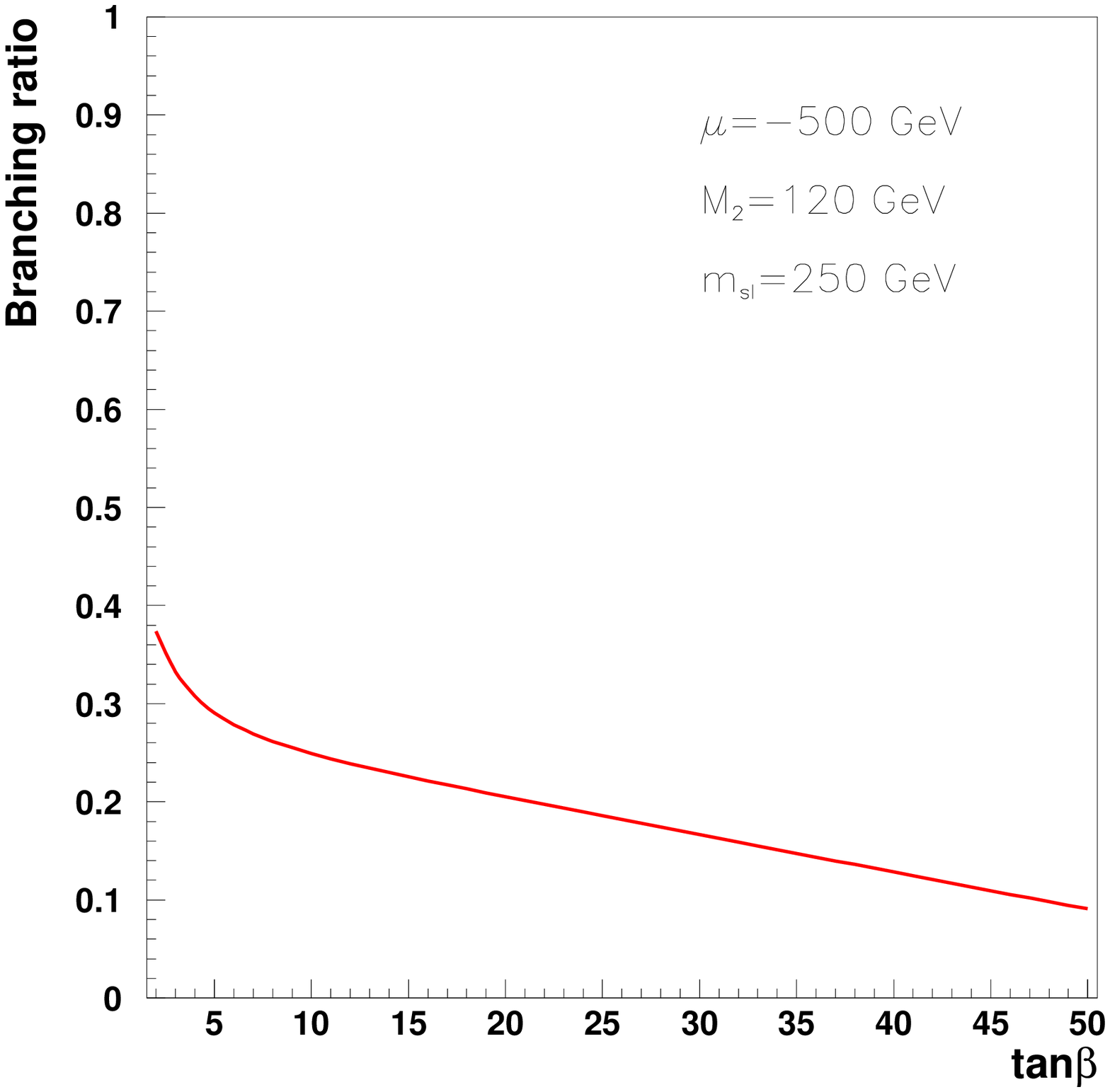}} &
  \resizebox{\linewidth}{!}{\includegraphics{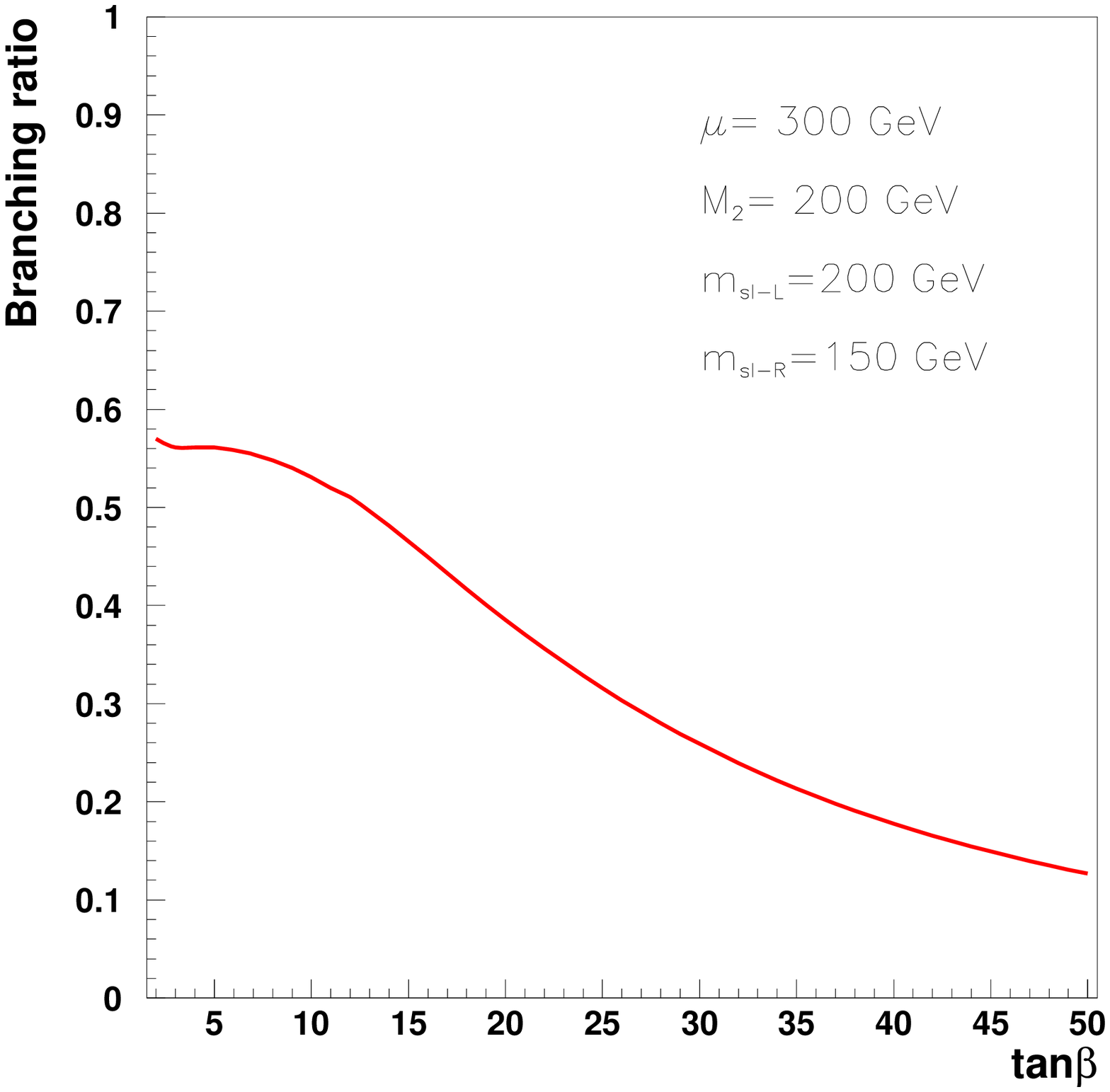}} \\
  \caption{Branching ratio of $\chi^0_2 \rightarrow l^+l^-\chi^0_1$ as a
  function of $\tan\beta$, in the case of $m_{\tilde{l}}$ $>$ $m_{\chi^0_2}$. ($l$ = $e$ + $\mu$)}
  \label{fig:nd1} &
  \caption{Branching ratio of $\chi^0_2 \rightarrow l^+l^-\chi^0_1$ as a
  function of $\tan\beta$, in the case of $m_{\tilde{l}}$ $<$ $m_{\chi^0_2}$. ($l$ = $e$ + $\mu$)} 
  \label{fig:nd2} \\
\end{2figures}

In fig. \ref{fig:nd1} we show the BR as a function of $\tan\beta$. Sleptons (including staus) 
are taken at 250 GeV and $\mu$ = -500 GeV. 
$\chi^0_2$ is then rather wino and $\chi^0_1$ is bino-dominated. 
Because of this, their coupling to the $Z^0$ is dynamically suppressed, and the $\chi^0_2
\rightarrow l^+ l^- \chi^0_1$ branching ratio will depend strongly on slepton masses.\\
If sleptons are lighter than the $\chi^0_2$, direct two-body decays of the neutralino into a slepton-lepton
pair are allowed, which may lead to large branching ratios.
In fig. \ref{fig:nd2} the evolution of the BR with $\tan\beta$ is shown for $M_2$ = 200 GeV,
$\mu$ = 300 GeV, $m_{\tilde{l_L}}$ = 200 GeV, $m_{\tilde{l_R}}$ = 150 GeV. This is however only valid in a rather limited region of the MSSM
parameter space, since often sneutrinos will be lighter than sleptons, causing the neutralinos to decay
purely into invisible particles. 
\\
The fall of BR($\chi^0_2 \rightarrow l^+l^-\chi^0_1$) with 
$\tan \beta$ in fig. \ref{fig:nd2} is compensated by a rise in BR($\chi^0_2 \rightarrow \tau^+ \tau^- \chi^0_1$).
This means that allowing taus in the final state could possibly extend our discovery reach towards 
higher $\tan\beta$ values. 
However, taus decay into leptons ($e$, $\mu$) in only $\sim$35\% of the cases, whilst the 
hadronic decay modes have detection efficiencies of $\sim$30\% \cite{sasha}. This, together with the fact 
there are up to four taus in the final state, makes that there is only a limited hope for a large 
improvement by including taus in the final state, but a dedicated study is needed. 

\section{Event generation}
The signal events are generated with SPYTHIA \cite{spythia}. For low $\tan\beta$ values 
($\tan\beta \sim 1$), the gluon-gluon fusion mechanism $gg \rightarrow A^0,H^0$ dominates the production. 
Due to the large coupling of the Higgses to $b\bar{b}$, the associated production $gg \rightarrow b \bar{b} A^0,H^0$ dominates
for $\tan\beta$ $>$ 5 \cite{dicus}. The CP-odd Higgs is produced more than the CP-even one because the 
$A^0 b \bar{b}$ coupling is directly proportional to $\tan \beta$, whilst the $H^0 b \bar{b}$ coupling is proportional to $\frac{\cos\alpha}{\cos\beta}$. 
In the decoupling regime (i.e. high values of $m_A$), both couplings become equal.
Besides these two main processes, we also included the $WW/ZZ$ fusion and Higgsstrahlung processes.  

We scanned the $A^0, H^0 \rightarrow  \chi^0_2 \chi^0_2 \rightarrow 4 l$ cross section
in the ($m_A$, $\tan\beta$) plane for different values of $\mu$ and
$m_{\tilde{l}}$. $M_1$ and $M_2$ were initially
kept on 60 and 120 GeV respectively, and no direct decays of neutralinos in sleptons were allowed. 
In figs. \ref{fig:sc1} and \ref{fig:sc2}, the plot 
of $\sigma \times$ BR for $\mu$ = -500 GeV and $m_{\tilde{l}}$ = 250 GeV is shown.  
Values of $\tan\beta$ $\lesssim$ 30 - 40 and $m_A$ $\lesssim$ 400 - 500
GeV seem to be favoured. One also notices that the pseudoscalar Higgs gives much 
higher cross sections than the scalar one.
\begin{2figures}{h}
  \resizebox{\linewidth}{!}{\includegraphics{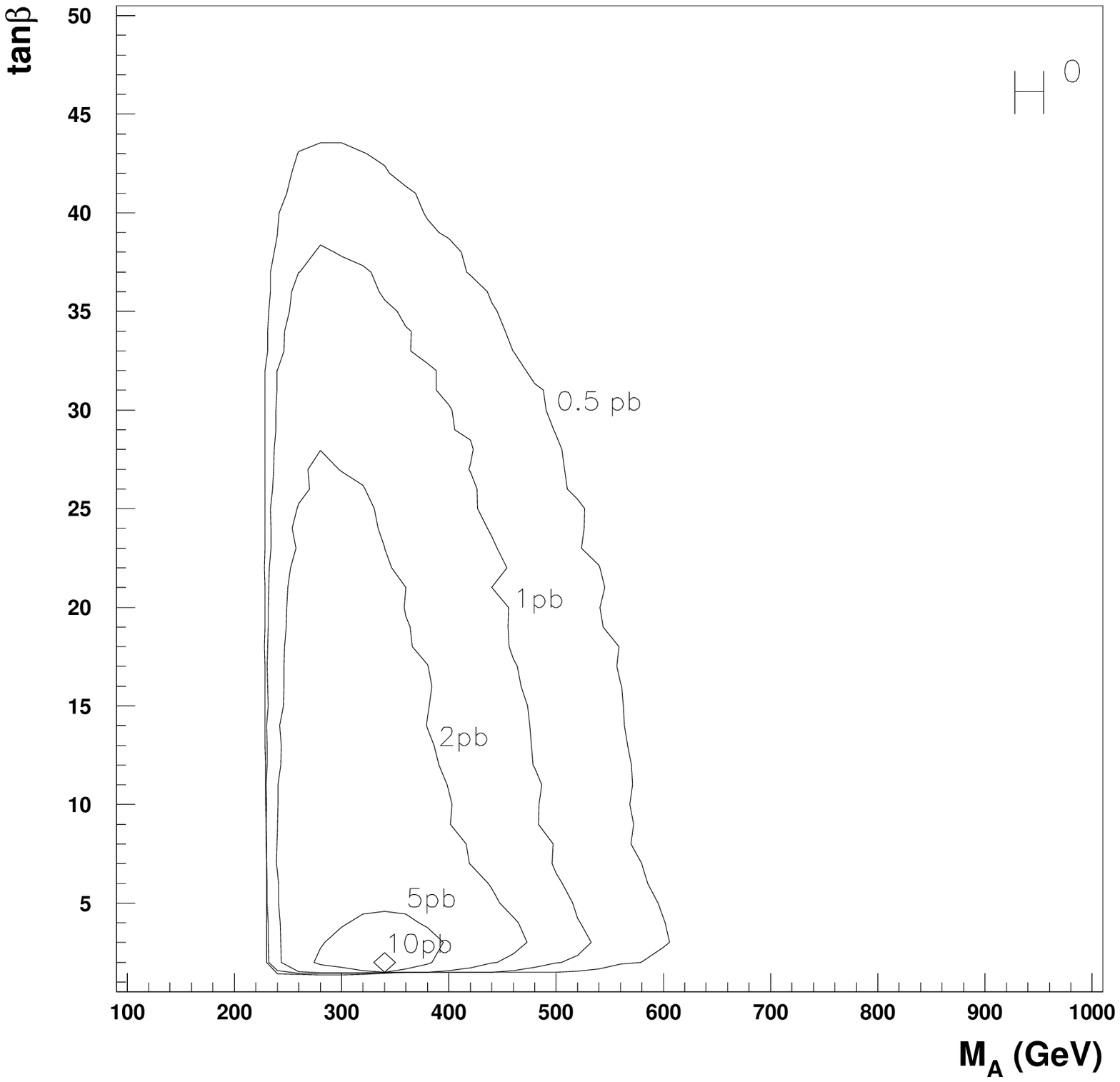}} &
  \resizebox{\linewidth}{!}{\includegraphics{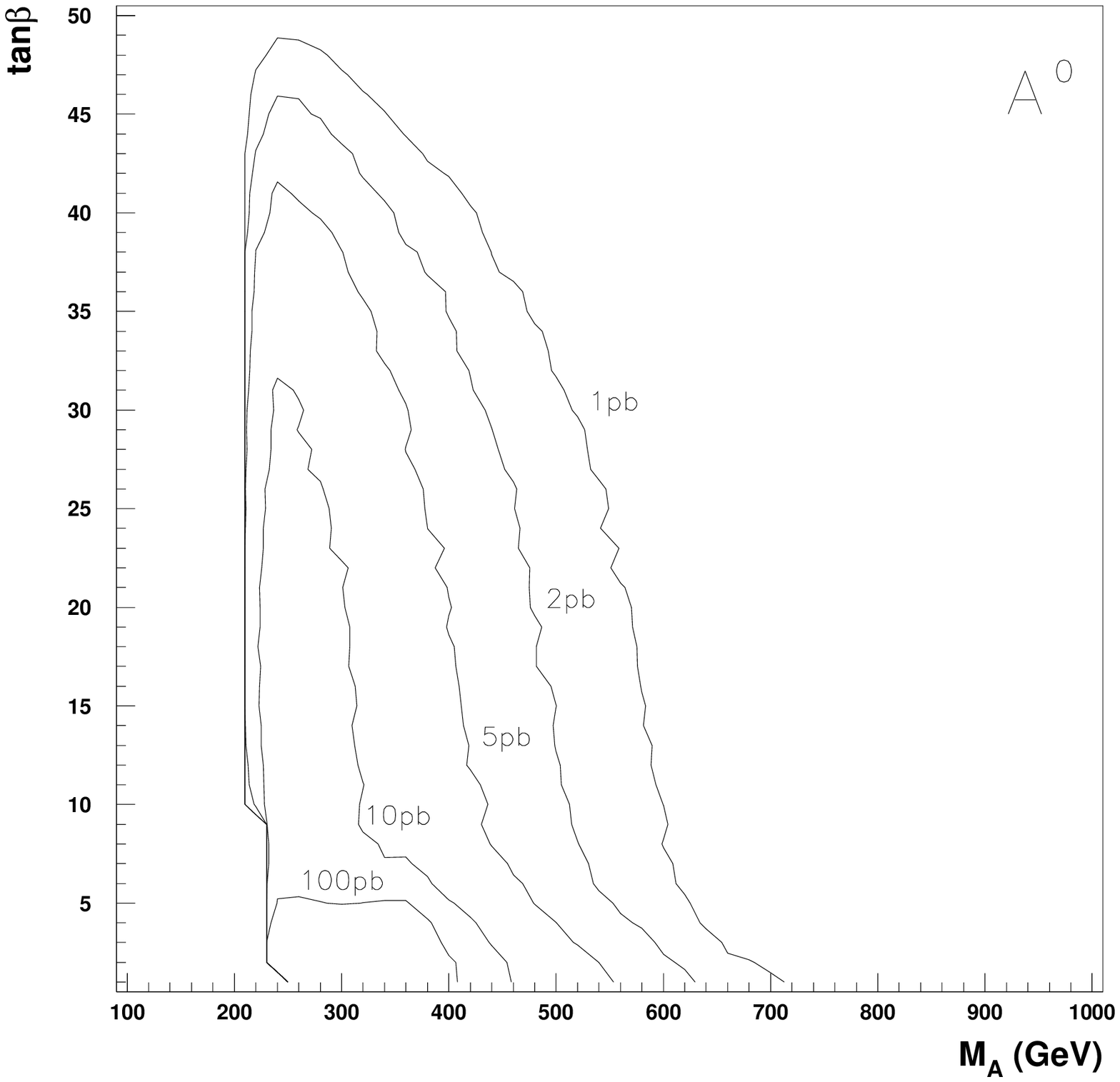}} \\
  \caption{$\sigma \times$ BR contours for $H^0$ in the $m_A-\tan\beta$ plane. Parameter values as described in the text.}
  \label{fig:sc1} &
  \caption{$\sigma \times$ BR contours for $A^0$ in the $m_A-\tan\beta$ plane. Parameter values as described in the text.}
  \label{fig:sc2} \\
\end{2figures}

For the background processes, PYTHIA 6.136 \cite{pythia} was used with a few bugs fixed both in the SUSY and general code.
The following SM backgrounds giving rise to four (real or fake) leptons in the final state have been generated:
$ZZ$, $ZW$, $Zb\bar{b}$, $Zc\bar{c}$, $Wt\bar{b}$ and $t\bar{t}$. Decays of $Z$ into $\tau$'s have been
included, since they might be dangerous due to their non-zero $E_T^{miss}$.    
For the SUSY backgrounds, we generated all pair production processes involving squarks,
gluinos, sleptons, charginos and neutralinos.\\
\\
The CMS detector response is simulated using the CMSJET fast Monte Carlo \cite{abdullin}.
The effects of pile-up at high luminosity running of LHC have not been included yet, but 
are expected to be minor in the four-lepton final state.

\section{Signal versus background discrimination}

In order to obtain a clear signal, we will have to 
discriminate between the signal events and background events
that contain a similar four lepton final state. Two categories of 
background have to be considered: Standard Model processes and SUSY backgrounds. \\
The main SM backgrounds are $ZZ$ and $t\bar{t}$ production. 
They are dangerous because of their large cross sections at the LHC.
In order to distinguish between events coming from the signal and from the SM background, 
we apply the following selection criteria:
\begin{itemize}
\item we require two pairs of isolated leptons with opposite sign and same flavour, with a
$P_T$ larger than 10 GeV and within $|\eta|$ $<$ 2.4 . 
The isolation criterion demands that there are no
charged particles with $P_T$ $>$ 1.5 GeV in a cone of R = 0.3 rad around each
lepton track, and that the sum of the transverse energy in the crystal towers between 
R = 0.05 and R = 0.3 rad is smaller than 3 GeV.
\item all dilepton pairs of opposite sign and same flavour that have an invariant mass in the
range $m_Z \pm 10$ GeV are rejected (Z veto).
\end{itemize}
Demanding four tightly isolated leptons with a transverse momentum higher than 10 GeV
is a powerful requirement in fighting the $t\bar{t}$ and $Wt\bar{b}$ background. 
An explicit $Z$ veto eliminates the $ZZ$, $WZ$, $Zb\bar{b}$ production and all other backgrounds
containing a $Z$ boson. Furthermore, to reduce 
$ZZ$ we also require a minimal missing transverse energy of 20 GeV. 
$ZZ$ events where one of the $Z$'s decays into taus, with the taus decaying
leptonically, can however pass this criterion.\\
\\
The SUSY background is more complex. 
Squark/gluino production is characterised by a large jet multiplicity (5 jets on average), a significant
$E_T^{miss}$ ($\gtrsim$ 100 GeV) and jet transverse momenta that are large compared to the
expectations for the signal. Selecting events with few, rather soft jets (e.g. $\le$ 
2 jets, $E_T$ of the hardest jet below 100 GeV) and with $E_T^{miss}$ $<$ 130 GeV allows us to eliminate most 
of these events. The $E_T^{jet}$ threshold can be lowered to 50 GeV if necessary. The squark/gluino - gaugino 
associated production can be eliminated this way too.
If we assume $m_{\tilde{q}, \tilde{g}}$ = 1000 GeV as in our default scenario, no squark/gluino events 
will survive the selection. 
In paragraph 7, the effects of lighter masses will be discussed.
\\
Slepton-slepton production predominantly ends up in a 2-lepton final state.
Sneutrino-sneutrino production remains however as the dominant SUSY background. 
It could possibly be distinguished from the signal because
of larger $E_T^{miss}$ and larger $p_T$ of the leptons,
as sneutrinos either decay into $\chi_2^0$ + $\nu$ (leading to
extra $E_T^{miss}$) or into $\chi_1^+$ + $l^-$ (leading to harder leptons).\\
Pair production of heavier neutralinos and charginos will lead to more and 
harder jets and will often contain $Z$ bosons in the final state. Direct
$\chi^0_2$-$\chi^0_2$ production gives the same signature as the signal,
but the production cross section is much smaller due to the strongly suppressed 
coupling of gauginos to the $Z$/$\gamma$ intermediate state. \\
\\
In figures \ref{fig:sel7} - \ref{fig:sel2}, the distributions of the
different kinematical variables for the signal and the total background (SM + SUSY) are plotted.
The parameters of the considered case are: $m_A$ = 350 GeV, $\tan\beta$ = 5, $M_1$ = 60 GeV, $M_2$ =
120 GeV, $\mu$ = -500 GeV, $m_{\tilde{l}}$ = 250 GeV, $m_{\tilde{q}, \tilde{g}}$ = 1000 GeV.
The dark shaded (blue) area is the part of the spectrum that is retained in the event selection.
\begin{2figures}{hp}
  \resizebox{\linewidth}{1.2\linewidth}{\includegraphics{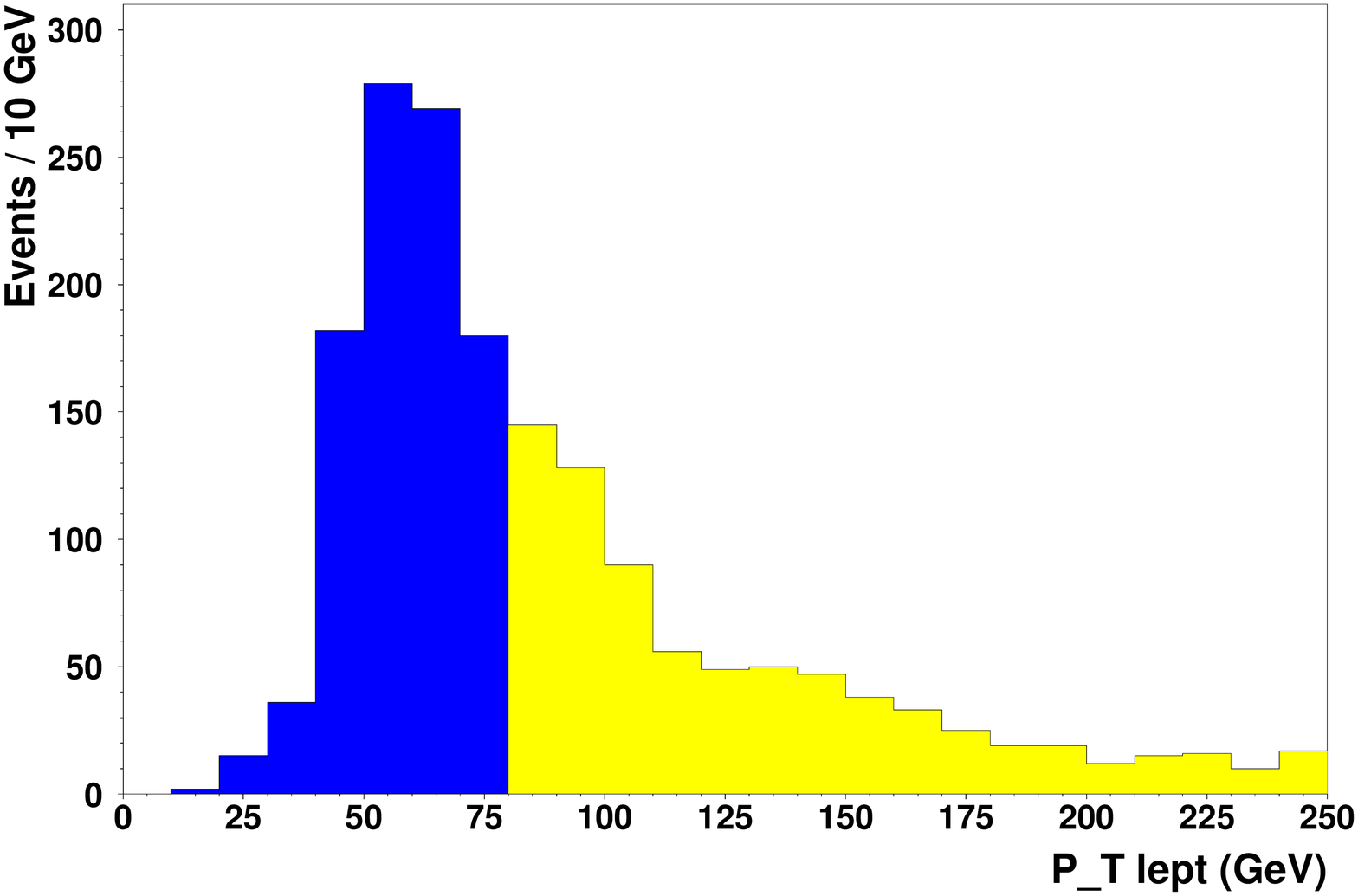}} &
  \resizebox{\linewidth}{1.2\linewidth}{\includegraphics{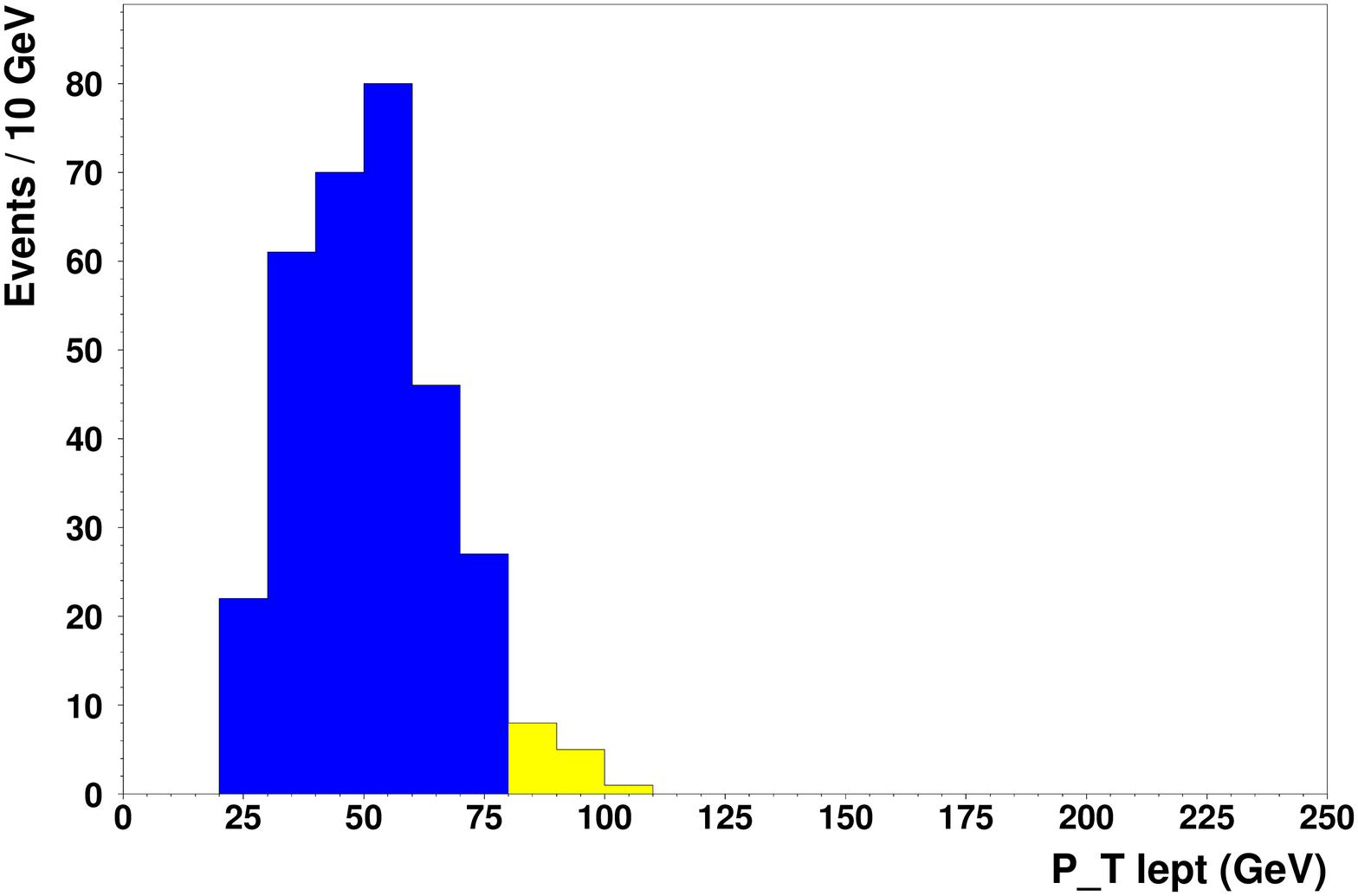}} \\
  \caption{$P_T$ spectrum of the hardest lepton in the event for the total background (SM + SUSY) at 100 $fb^{-1}$.}
  \label{fig:sel7} &
  \caption{$P_T$ spectrum of the hardest lepton in the event for the signal at 100 $fb^{-1}$.}
  \label{fig:sel8} \\
\end{2figures}
\begin{2figures}{hp}
  \resizebox{\linewidth}{1.2\linewidth}{\includegraphics{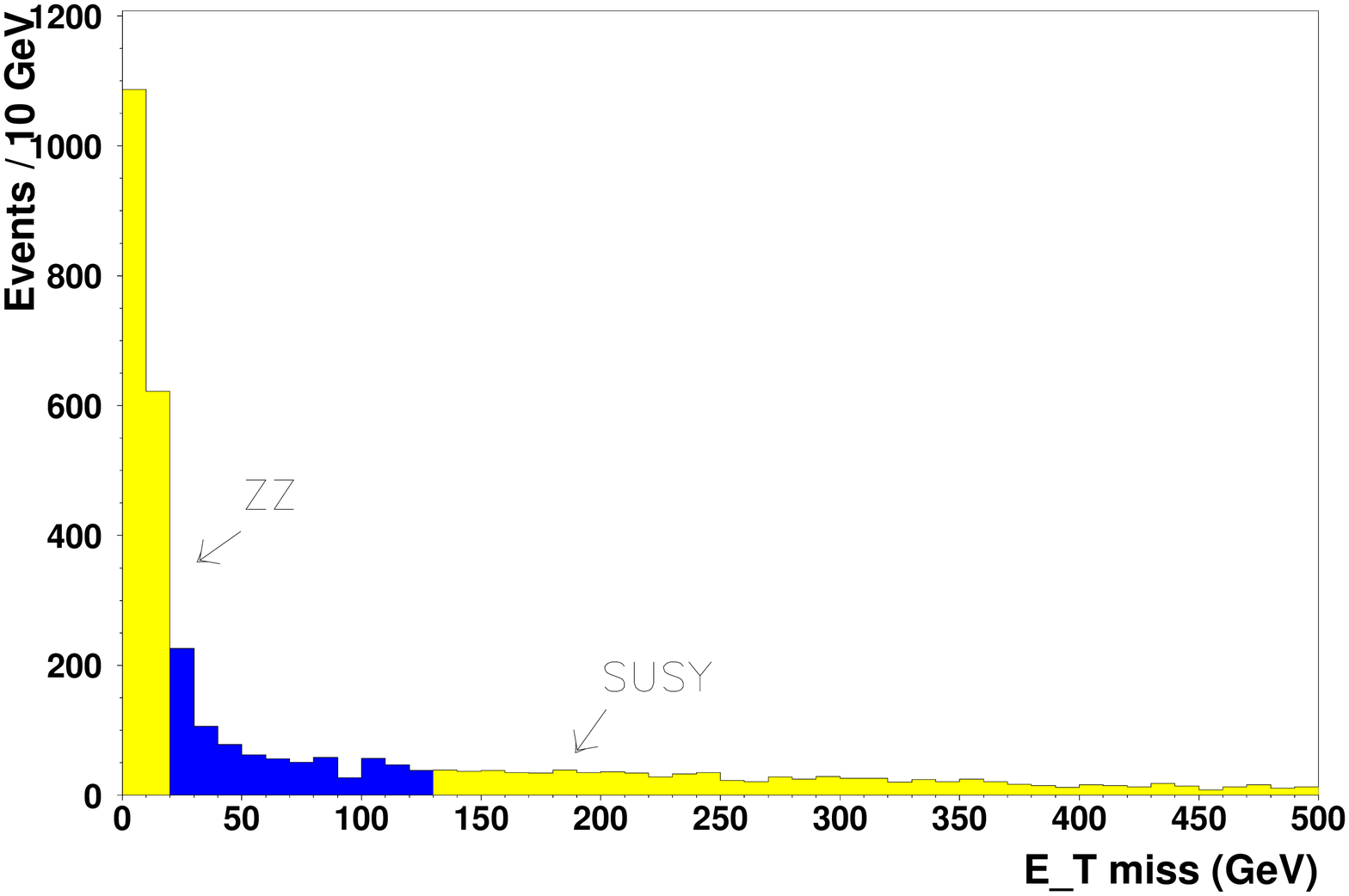}} &
  \resizebox{\linewidth}{1.2\linewidth}{\includegraphics{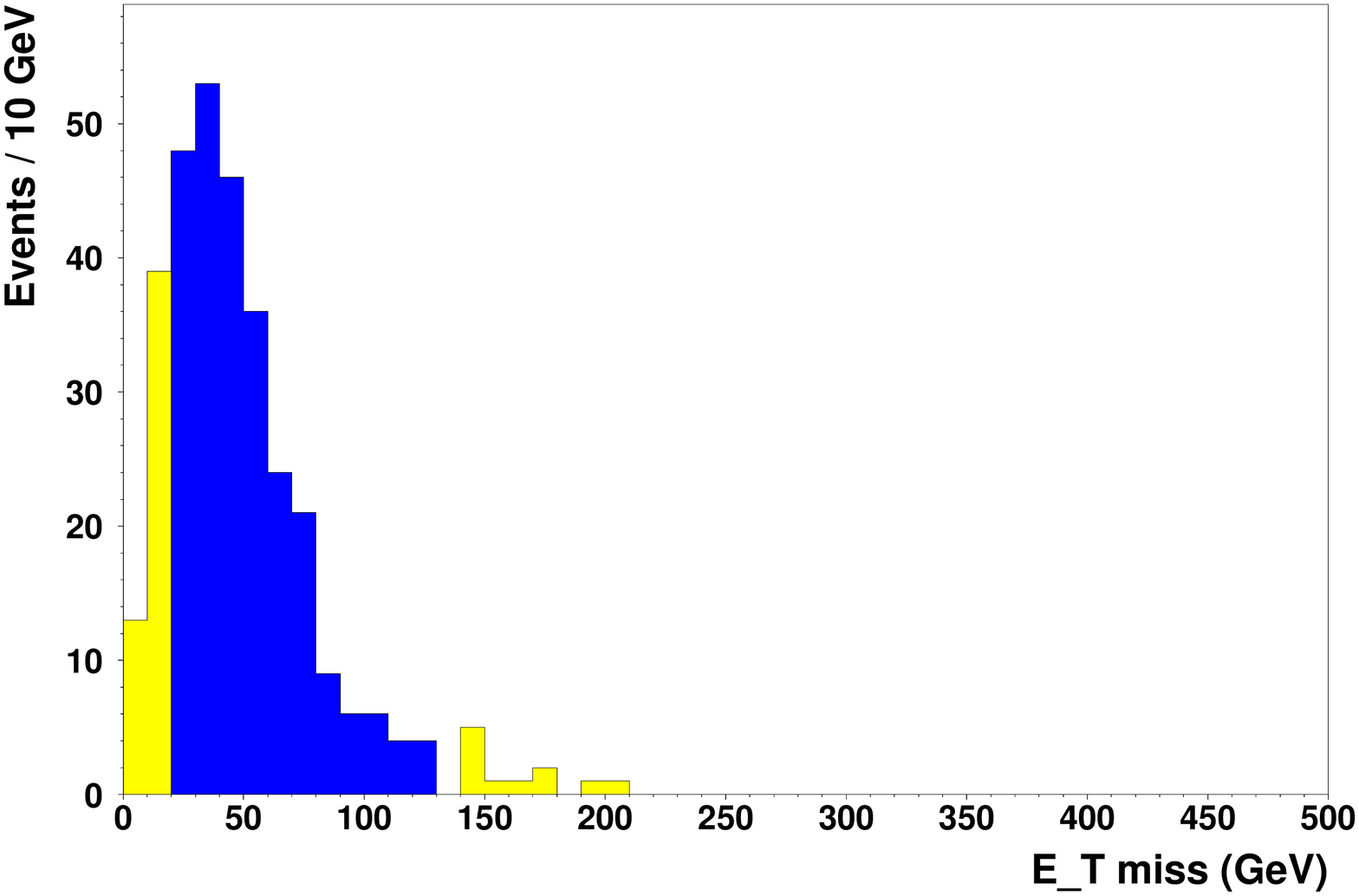}} \\
  \caption{Missing $E_T$ distribution for the total background (SM + SUSY) at 100 $fb^{-1}$.}
  \label{fig:sel5} &
  \caption{Missing $E_T$ distribution for the signal at 100 $fb^{-1}$.}
  \label{fig:sel6} \\
\end{2figures}
\begin{2figures}{hp}
  \resizebox{\linewidth}{1.2\linewidth}{\includegraphics{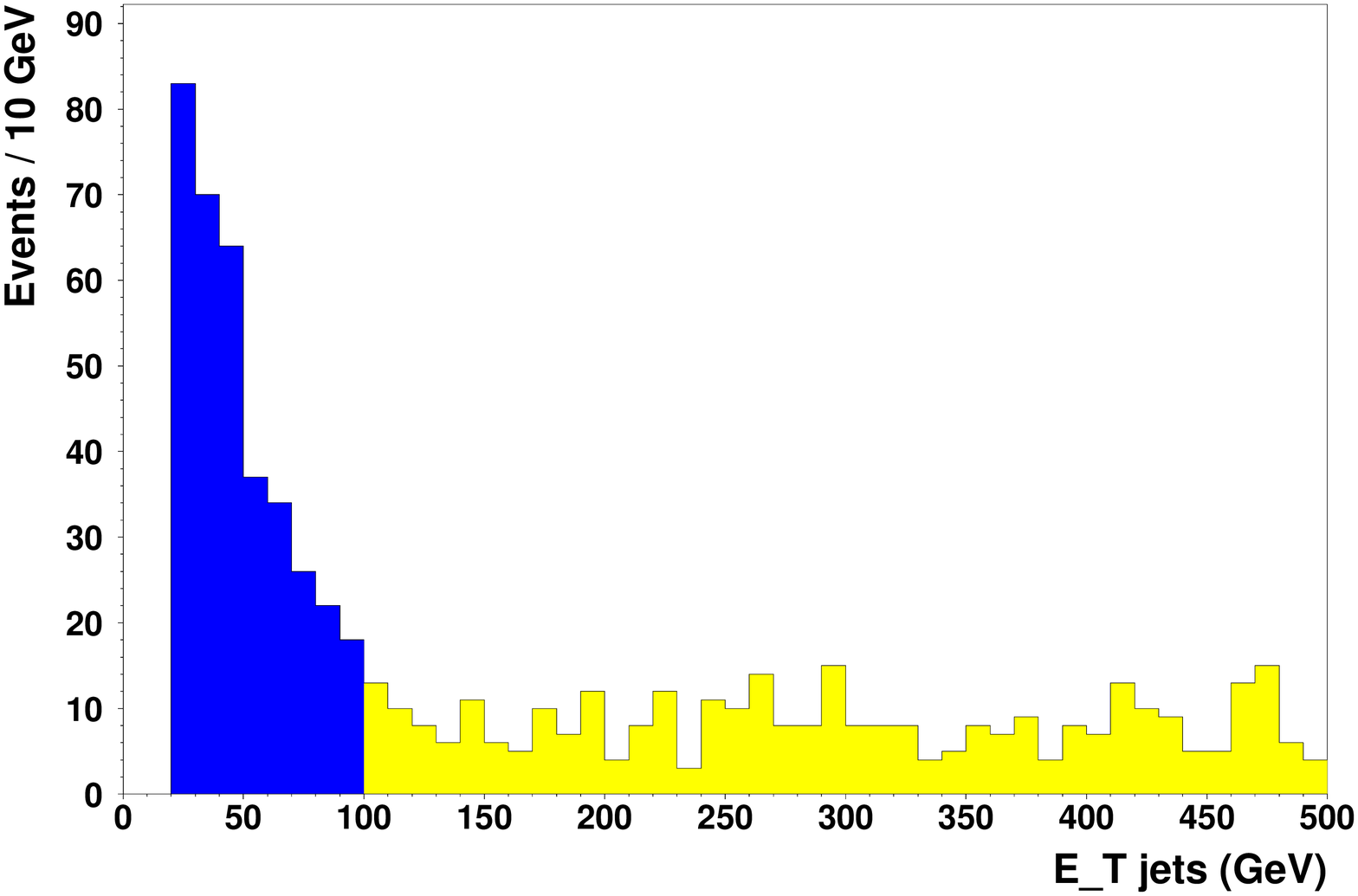}} &
  \resizebox{\linewidth}{1.2\linewidth}{\includegraphics{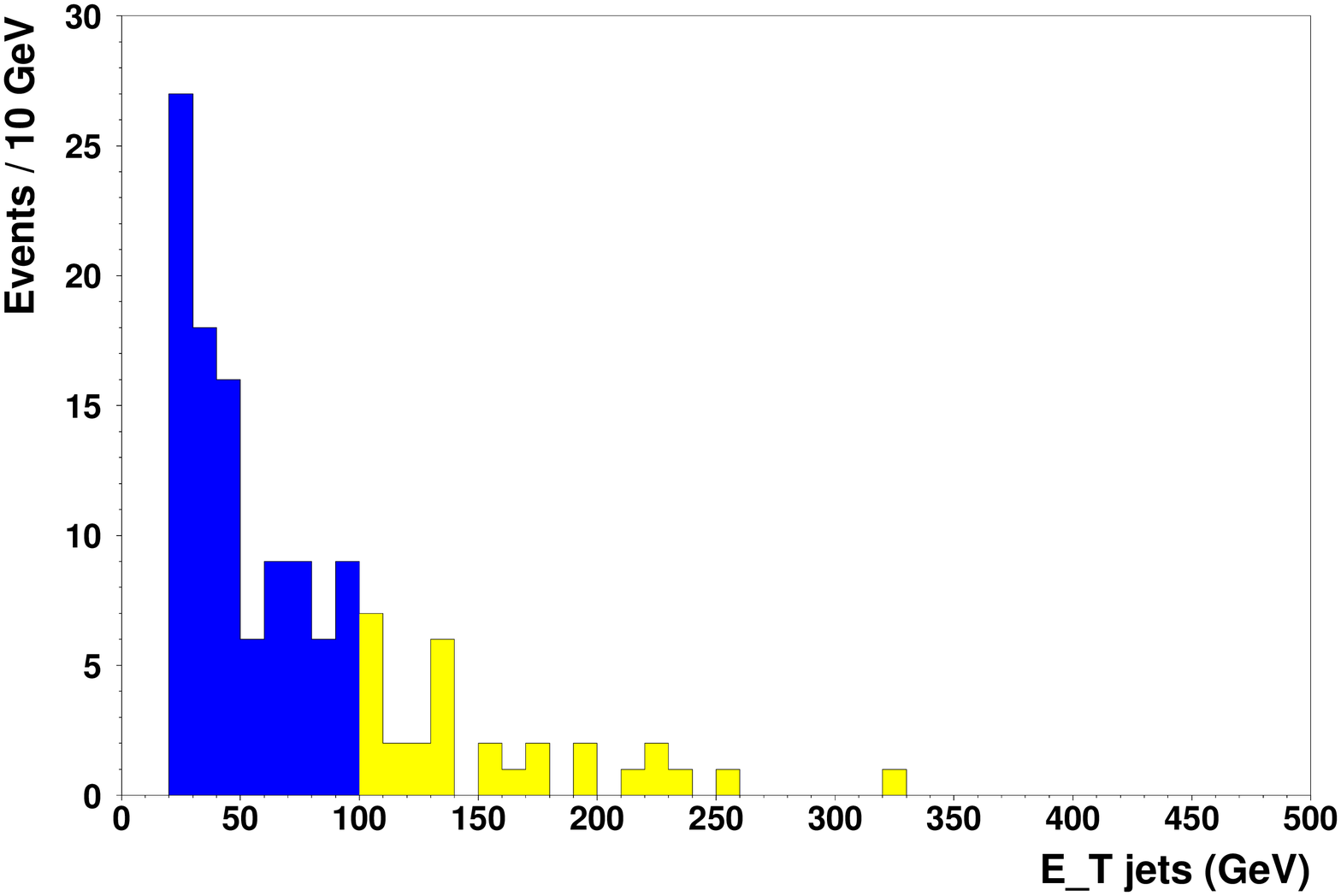}} \\
  \caption{Transverse energy distribution of the hardest jet in the event for the total background (SM + SUSY) at 100 $fb^{-1}$.}
  \label{fig:sel3} &
  \caption{Transverse energy distribution of the hardest jet in the event for the signal at 100 $fb^{-1}$.}
  \label{fig:sel4} \\
\end{2figures}
\begin{2figures}{hp}
  \resizebox{\linewidth}{1.2\linewidth}{\includegraphics{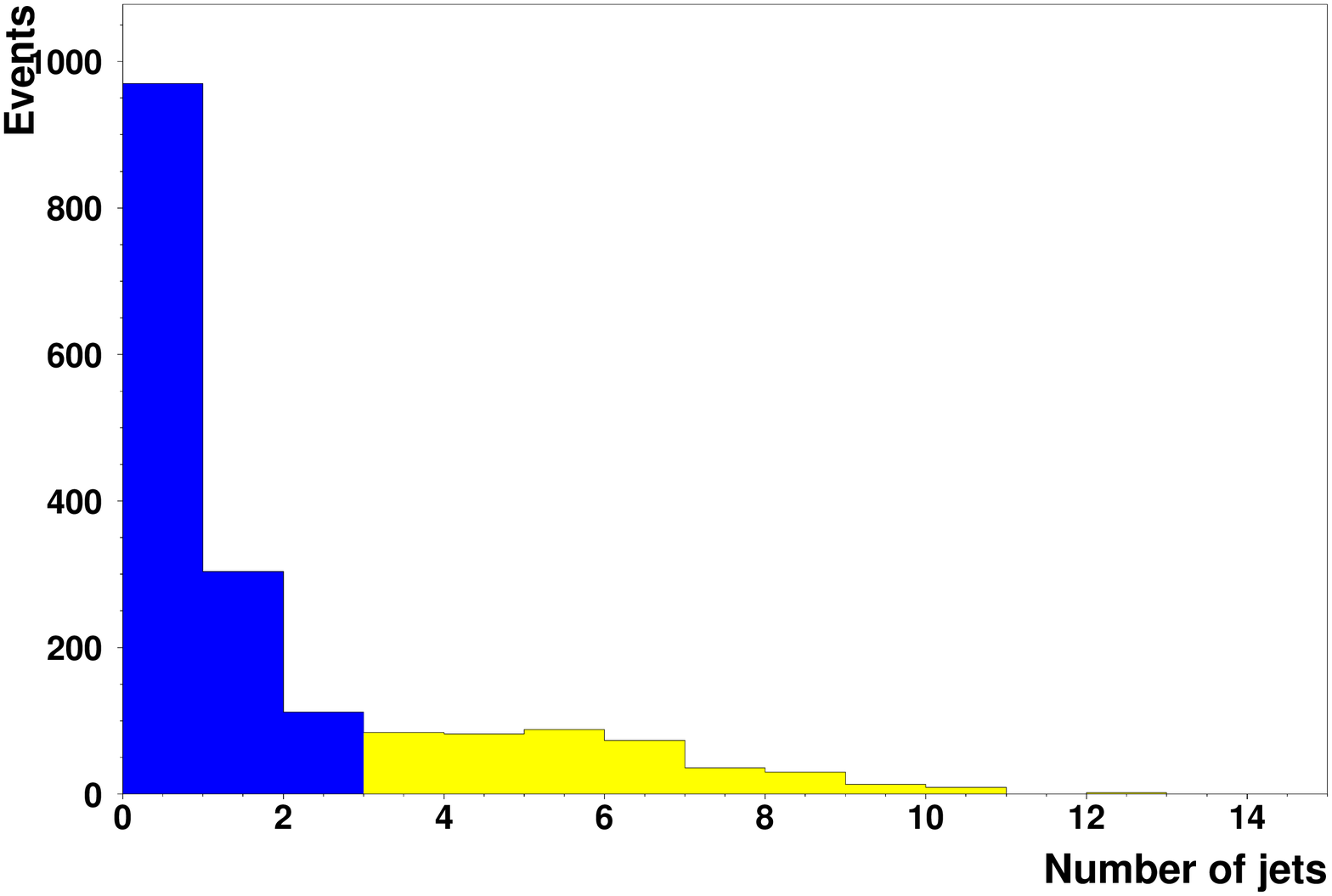}} &
  \resizebox{\linewidth}{1.2\linewidth}{\includegraphics{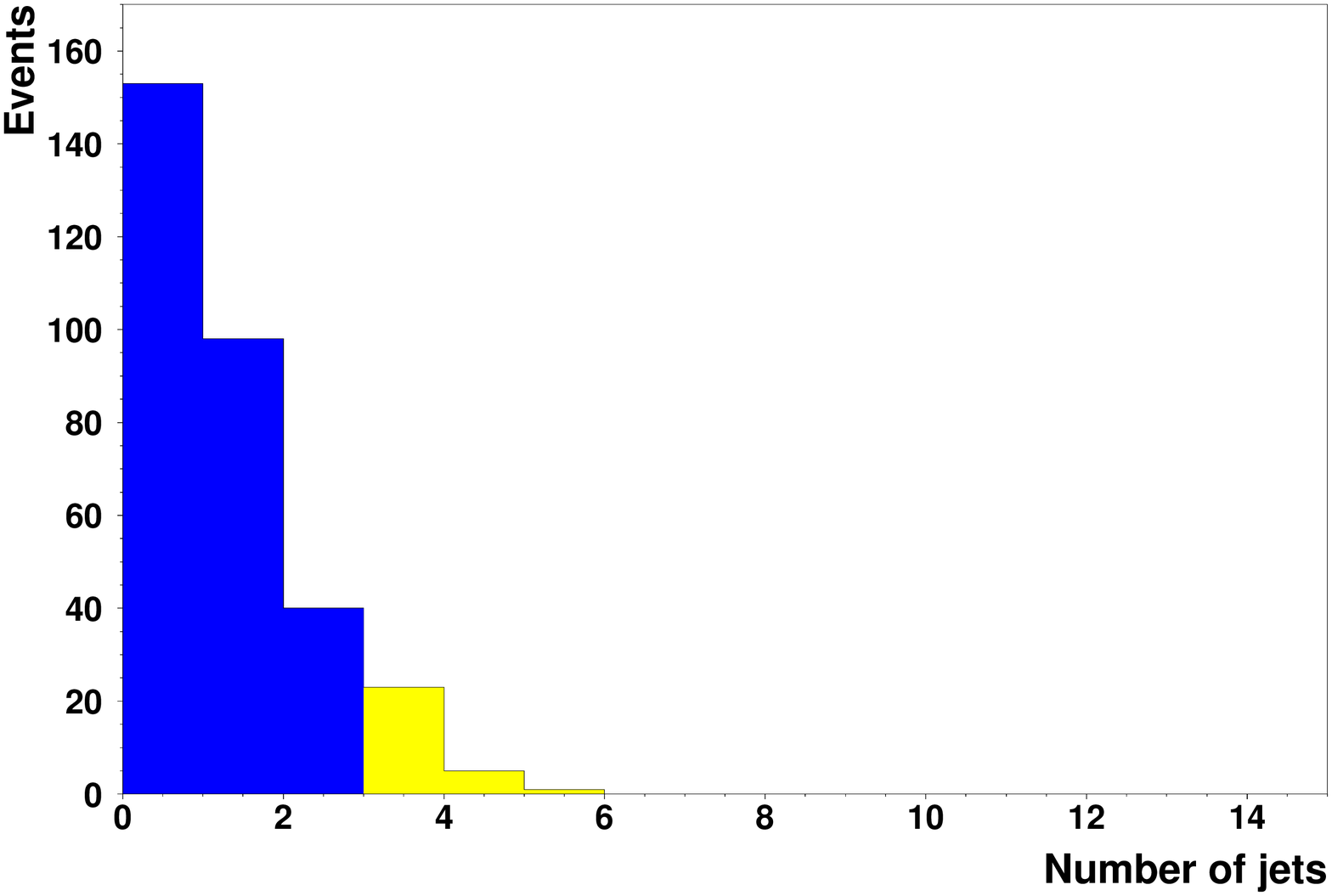}} \\
  \caption{Jet multiplicity for the total background (SM + SUSY) at 100 $fb^{-1}$.}
  \label{fig:sel1} &
  \caption{Jet multiplicity for the signal at 100 $fb^{-1}$.}
  \label{fig:sel2} \\
\end{2figures}
\begin{2figures}{htp}
  \resizebox{\linewidth}{1.2\linewidth}{\includegraphics{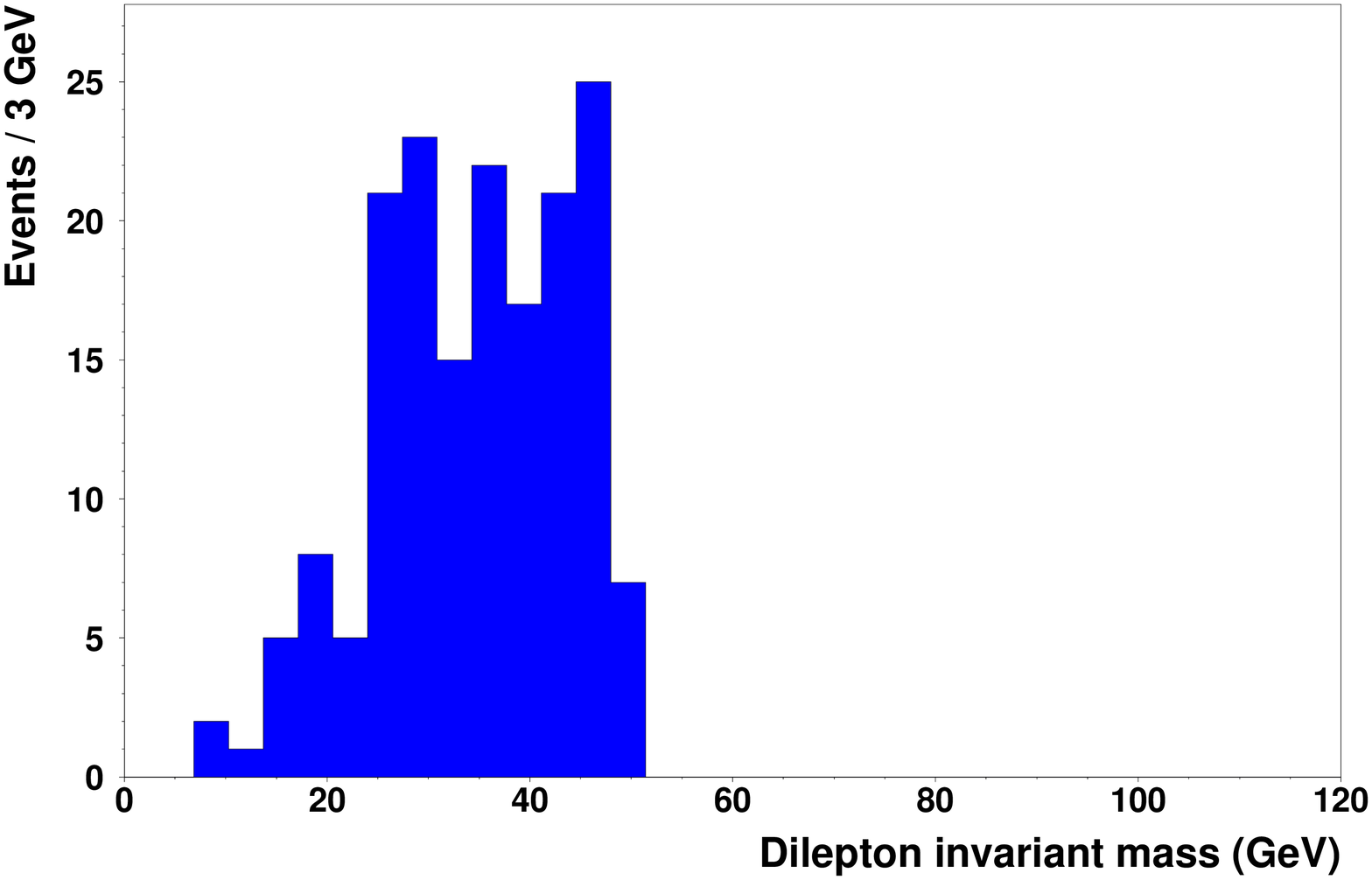}} &
  \resizebox{\linewidth}{1.2\linewidth}{\includegraphics{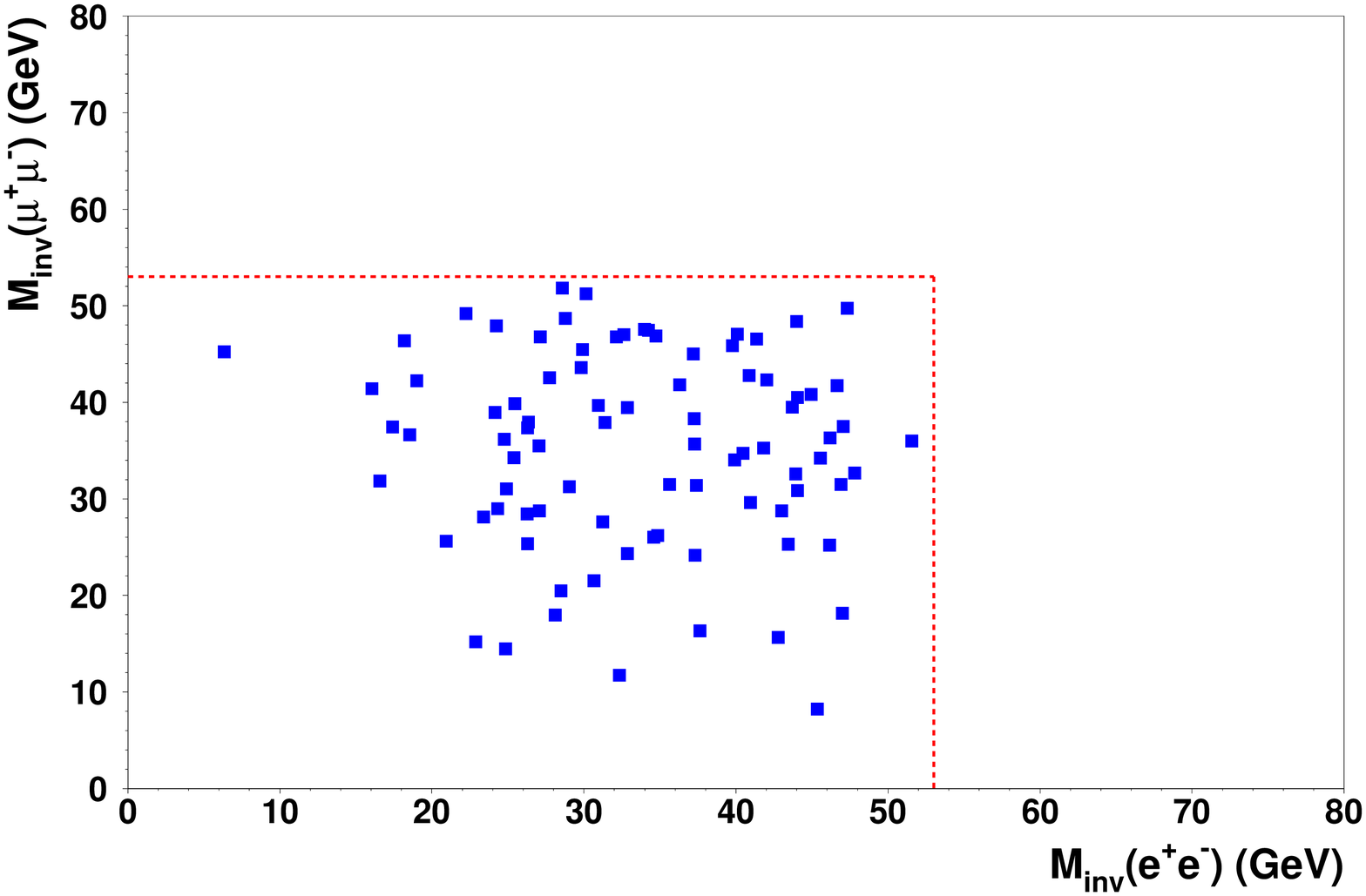}} \\
  \caption{Kinematical edge in the dilepton (opposite sign - same flavour) invariant mass distribution.}
  \label{fig:edge1} &
  \caption{Double kinematical edge in the di-electron versus dimuon invariant mass.}
  \label{fig:edge2} \\
\end{2figures}
\\
\noindent
In view of these distributions, we will apply the following search strategy:
events are selected with $E_T^{miss}$ smaller than 130 GeV (to suppress the SUSY background), 
but larger than 20 GeV (to suppress ZZ background). The $P_T$ of the hardest lepton should be less than 80
GeV. The $E_T$ of the harderst jet in the event is taken smaller than 100 GeV. 
In addition, we could also make a jet multiplicity requirement ($\le$ 2 jets), but this seems to be needed only
if squarks and/or gluinos would be light (cfr. paragraph 7). 
The four lepton invariant mass of the signal events should not exceed $m_A$ - $2m_{\chi^0_1}$. If the mass of the
lightest neutralino is approximately known at the time of the analysis, one could set a limit at $m_{llll}$ $\leq$ 230 GeV. \\
The number of signal and background events remaining after applying this selection  
is, for the considered case, given in table 1. 
\begin{table}[h]\label{tab:1}
    \caption{Number of events after successive cuts (at 100 $fb^{-1}$). As parameters  
    were used: $m_A$ = 350 GeV, $\tan\beta$ = 5, $M_1$ = 60 GeV, $M_2$ =
120 GeV, $\mu$ = -500 GeV, $m_{\tilde{l}}$ = 250 GeV, $m_{\tilde{q}, \tilde{g}}$ = 1000 GeV.}
    \begin{center}
    \begin{tabular}{|l||c|c|c|c|c|c|} \hline 
        Process           & 4l events (isol.)& Z-veto & $P_T^{lept}$ cut & $E_T^{miss}$ cut& $E_T^{jet}$ cut & 4l inv. mass \\ \hline
    $\tilde{q}$, $\tilde{g}$  &    421    &   206  &     60        &      8      &      1   &    1          \\ 
    $\tilde{t} \tilde{t}$   &      10    &     4  &      2         &      0      &      0   &    0          \\ 
    $\tilde{l}$,$\tilde{\nu}$ &    191    &    92  &     20        &     15      &     15    &   11          \\ 
    $\tilde{q} \tilde{\chi}$&     41    &    23  &      7          &      0      &      0      & 0           \\
    $\tilde{\chi} \tilde{\chi}$&     40    &    20  &      13     &      5       &      4   &  3             \\ \hline
    total SUSY bkg.       &    703    &   345  &     102         &     28        &     20      &  15           \\
    $ZZ$                  &   2106    &    80  &      79         &     10        &     10      &   2        \\ \hline    
    total bkg.            &   2809    &   425  &    181          &     38        &     30      &  17         \\ \hline 
    $H,A$ signal          &    268    &   232  &    218          &    179        &    164      &  164       \\ \hline
      \end{tabular}
    \end{center}
\end{table}
\\
An extra feature that can be exploited in the signal versus background discrimination is the shape of 
the dilepton invariant mass spectrum. For signal events both $\chi^0_2 \rightarrow l^+l^-\chi^0_1$ 
decays will result in a kinematical edge at $m_{\chi^0_2}$ - $m_{\chi^0_1}$ in the dilepton invariant mass.
In figs. \ref{fig:edge1} and \ref{fig:edge2}, this edge can be seen for our case
where $m_{\chi^0_2}$ - $m_{\chi^0_1}$ $\approx$ 50 GeV. To avoid improper
combinations of leptons, one can use selected events 
where one neutralino decays to electrons and the other one to muons.
Selecting events in which both dilepton pairs are close to the kinematical edge may allow for
the direct reconstruction of the $A^0$ / $H^0$ mass, if the statistics are sufficient.
\begin{2figures}{ht}
  \resizebox{\linewidth}{\linewidth}{\includegraphics{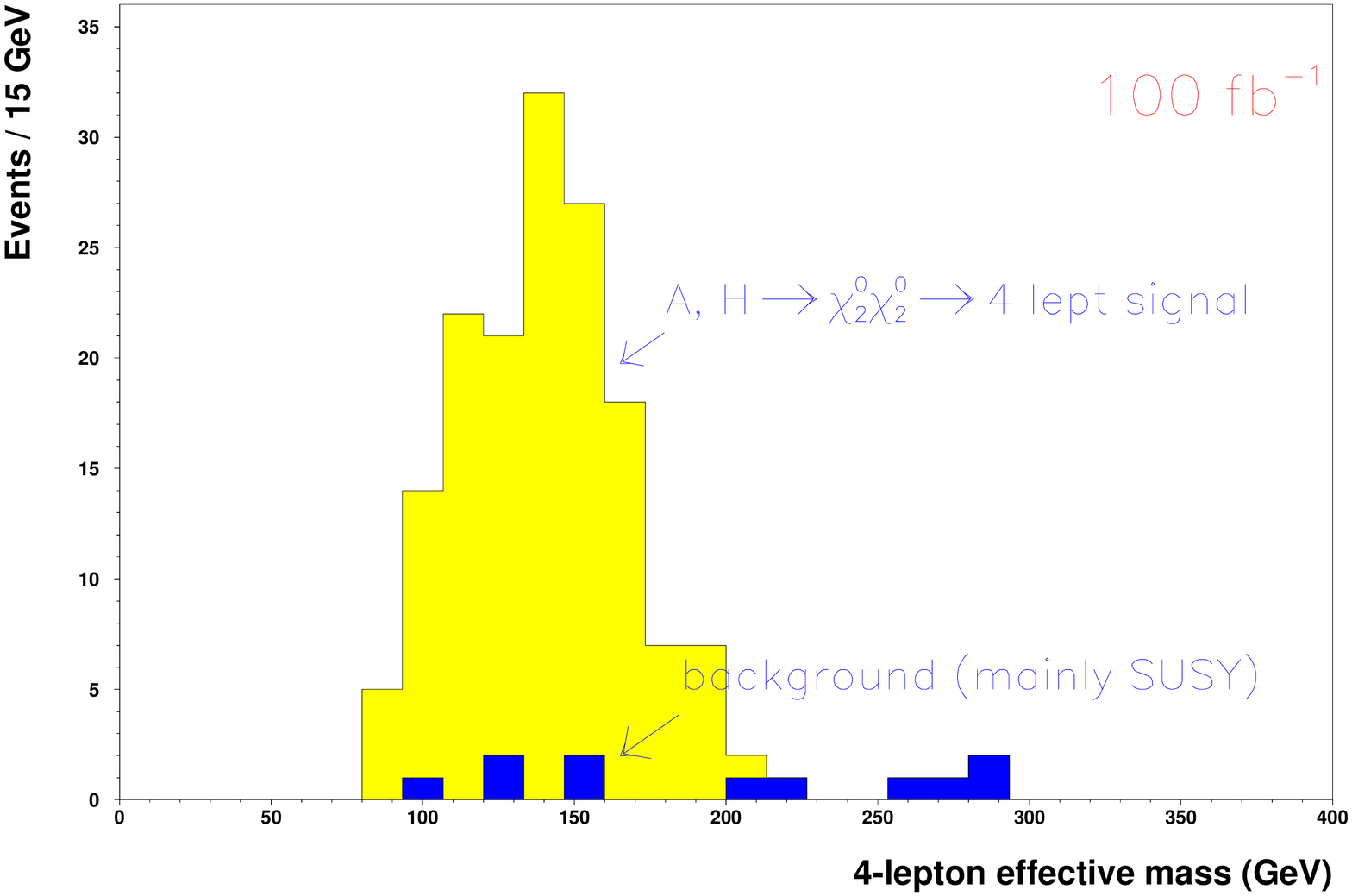}} &
  \resizebox{\linewidth}{\linewidth}{\includegraphics{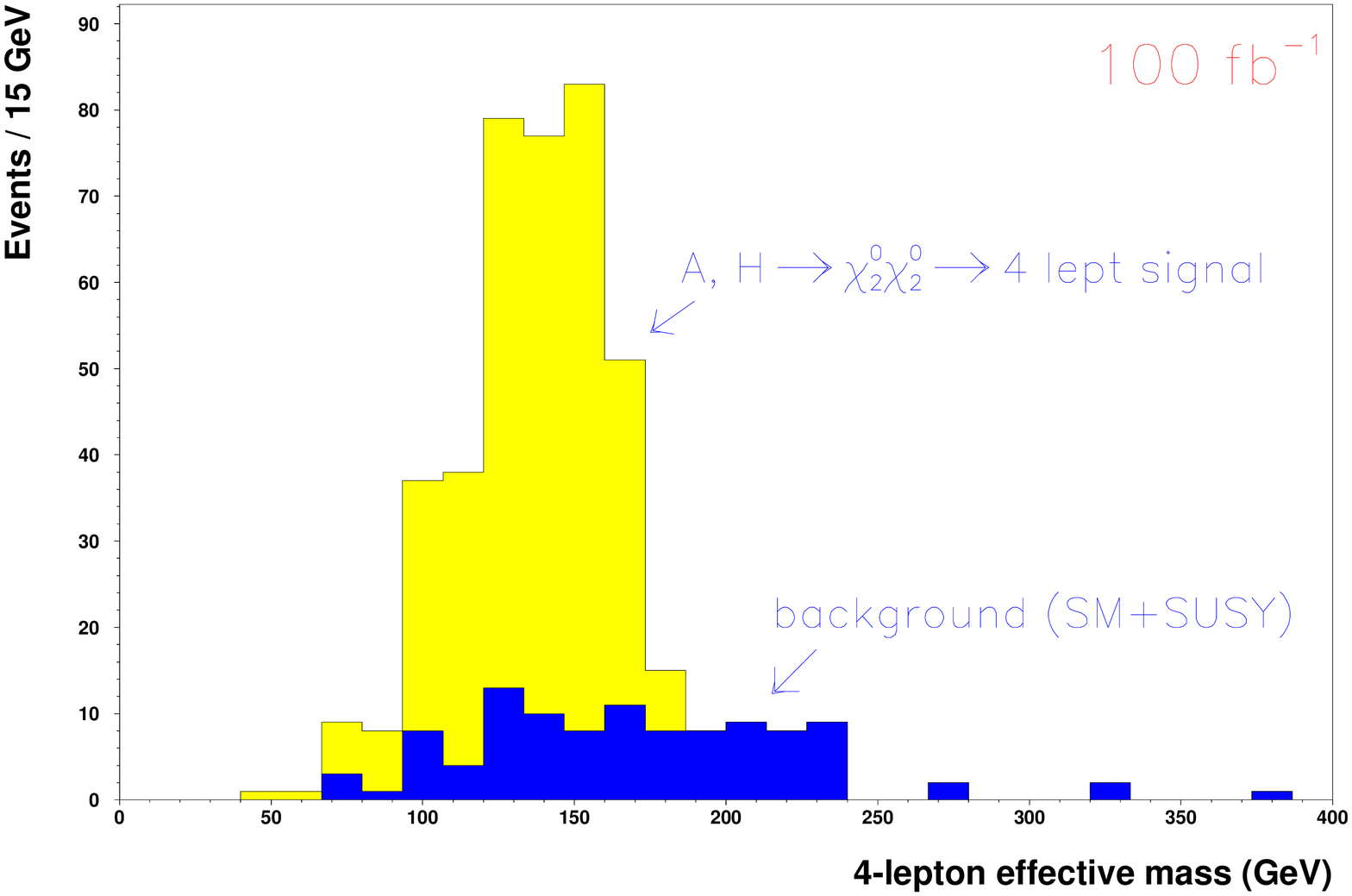}} \\
 \caption{Four lepton invariant mass for signal versus background (SM + SUSY) for point A: $m_A$ = 350
  GeV, $\tan\beta$ = 5, $M_1$ = 60 GeV, $M_2$ = 120 GeV, $\mu$~= -500 GeV, $m_{\tilde{l}}$ = 250 GeV, 
  $m_{\tilde{q}, \tilde{g}}$ = 1000 GeV (100 $fb^{-1}$).} 
  \label{fig:sel0a}&
 \caption{Four lepton invariant mass for signal versus background (SM + SUSY) for point B: $m_A$ = 350
  GeV, $\tan\beta$ = 10, $M_1$ = 100 GeV, $M_2$ = 200 GeV, $\mu$~= 350 GeV, $m_{\tilde{l_L}}$ = 200 GeV, 
  $m_{\tilde{l_R}}$ = 150 GeV, $m_{\tilde{q}, \tilde{g}}$ = 550 GeV (100 $fb^{-1}$).}
  \label{fig:sel0b} \\
\end{2figures}

After this selection, we can plot the four lepton invariant mass of the events.
In fig. \ref{fig:sel0a}, we show the case of $m_A$ = 350 GeV and $\tan\beta$ = 5, $M_1$ = 60 GeV, $M_2$ =
120 GeV, $\mu$ = -500 GeV, $m_{\tilde{l}}$ = 250 GeV, $m_{\tilde{q}, \tilde{g}}$ = 1000 GeV (point A).
In fig. \ref{fig:sel0b}, we present a case where direct decays of neutralinos into right-handed sleptons 
is allowed: $m_A$ = 350 GeV, $\tan\beta$ = 10, $M_1$ = 100 GeV, $M_2$ = 200 GeV,
$\mu$ = 350 GeV, $M_{\tilde{l_L}}$ = 200 GeV, $M_{\tilde{l_R}}$ = 150 GeV, $M_{\tilde{q}, \tilde{g}}$ = 550 GeV (point B).
At these chosen points in parameter space, the signal entirely dominates the 4-lepton sample, with less 
than 10 percent of background events.\\

\section{Discovery reach in the $m_A$ - $\tan\beta$ plane}
\begin{figure}[ht]
  \begin{center}
  \resizebox{\linewidth}{\linewidth}{\includegraphics{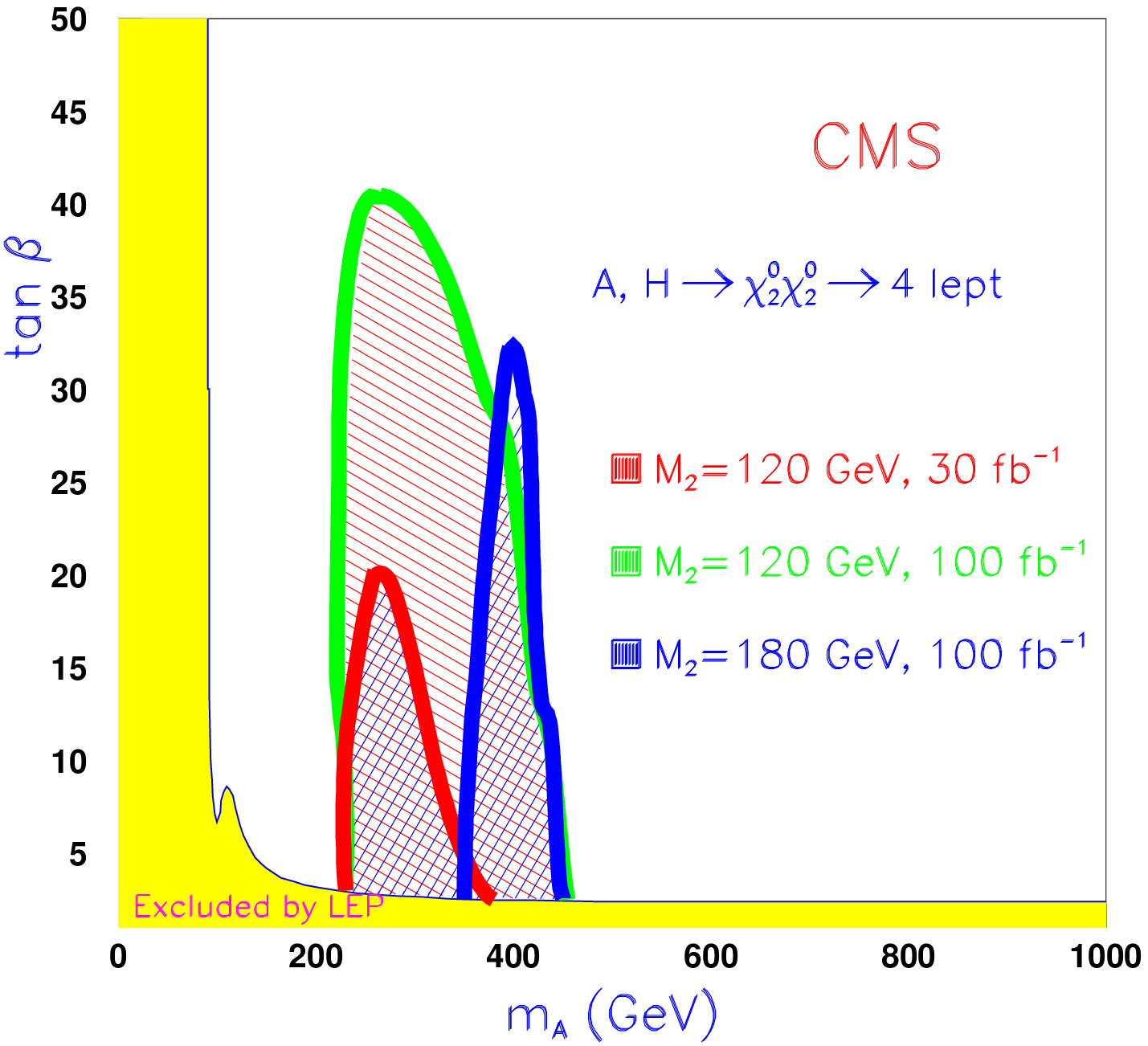}}
  \vspace{1.5cm}
  \caption{5$\sigma$-discovery contours in the $m_A-\tan\beta$ plane for 30 and 100
  $fb^{-1}$ and for $M_2$ = 120 and 180 GeV.\hspace{8cm} The other MSSM parameters are $\mu$ = -500 GeV, 
  $M_1$ = 0.5*$M_2$, $m_{\tilde{l}}$ = 250 GeV, $m_{\tilde{q}, \tilde{g}}$ = 1000 GeV.}  
  \label{fig:dr1} 
  \end{center}
\end{figure}

As a next step, we investigated the domain of $m_A$ - $\tan\beta$ parameter space where
the $A/H \rightarrow \chi_2^0 \chi_2^0 \rightarrow 4 l + E_T^{miss}$ final states would be
detectable. 
We tried to optimize our selection criteria to get the best S/B ratio and best signal observability
by sliding upper and lower cuts on selection variables.\\
As a criterion for discovery, we require the significance $\sigma$ $\ge$ 5,
with $\sigma$ defined conservatively as $\sigma = S / \sqrt{S+B}$.
$S$ and $B$ are the number of signal and background events respectively.\\
In fig. \ref{fig:dr1} the region in the $m_A-\tan\beta$ plane is plotted where a $5\sigma$-discovery 
could be made for an integrated luminosity of 30 and 100 $fb^{-1}$, provided we understand
the nature of both SM and SUSY backgrounds in the final state topology. For this plot, 
two different $\chi^0_2$ masses were considered ($M_2$ = 120 and 180 GeV) and the
other MSSM parameters were chosen as $M_1$ = 0.5*$M_2$, $\mu$ = -500 GeV,
$m_{\tilde{l}}$ = 250 GeV, $m_{\tilde{q}, \tilde{g}}$ = 1000 GeV, so direct decays of neutralinos into sleptons
are not allowed. 
The discovery region starts where the $\chi^0_2\chi^0_2$ decay becomes kinematically 
accessible, $m_A \, \ge \, 2m_{\chi_2^0}$, this is $\approx$ 230 GeV for $M_2$ = 120 GeV;  
the upper reach in $m_A$ is, for high values of $|\mu|$, mainly determined by the $A^0,H^0$ production cross section, which drops
with $m_A$ as a power law. The reach in $\tan \beta$ is determined by the branching ratio of $A^0,H^0 \rightarrow \chi^0_2\chi^0_2$ and $\chi_2^0
\rightarrow l^+ l^- \chi_1^0$.\\
At 30 $fb^{-1}$, the discovery region reaches $m_A \approx $ 350 GeV and $\tan\beta \approx$ 20.
For 100 $fb^{-1}$, values of $\tan\beta$ $\approx$ 40 and masses up to $m_A \approx $ 450 GeV are accessible.
In the case that $\chi^0_2 \rightarrow \tilde{l}l$ decays are dominating, even higher values of $m_A$ may be observable, 
but the reach in $\tan\beta$ will be lower.\\
These areas are covering the otherwise difficult region (partly at 30 $fb^{-1}$ and fully at 
100 $fb^{-1}$) of MSSM parameter space that is not easily accessible for SM decays of SUSY Higgses
- except for the $h \rightarrow b\bar{b}$ mode.

\section{Dependence of the discovery potential on the other MSSM parameters}\label{chap:slep}
Since the discovery reach depends strongly on the nature of the $\chi^0_2 \chi^0_2$ 
intermediate state, we have to investigate the effect of varying 
- apart from $m_A$ and $\tan\beta$ - also the other important MSSM parameters 
$\mu$, $m_{\tilde{l}}$, $M_1$, $M_2$ and $m_{\tilde{g}, \tilde{q}}$.\\
There are two main aspects to investigate: the $\mu$ - $m_{\tilde{l}}$ dependence of the
discovery potential, which will be crucial for the signal; and the effect of 
light squarks and gluinos, leading to additional large SUSY backgrounds. 
The impact of the $M_1$ and $M_2$ parameters is discussed too.\\

{\bf{Discovery potential in the $\mu$ - $m_{\tilde{l}}$ plane}}

\begin{figure}[ht]
  \begin{center}
  \resizebox{\linewidth}{\linewidth}{\includegraphics{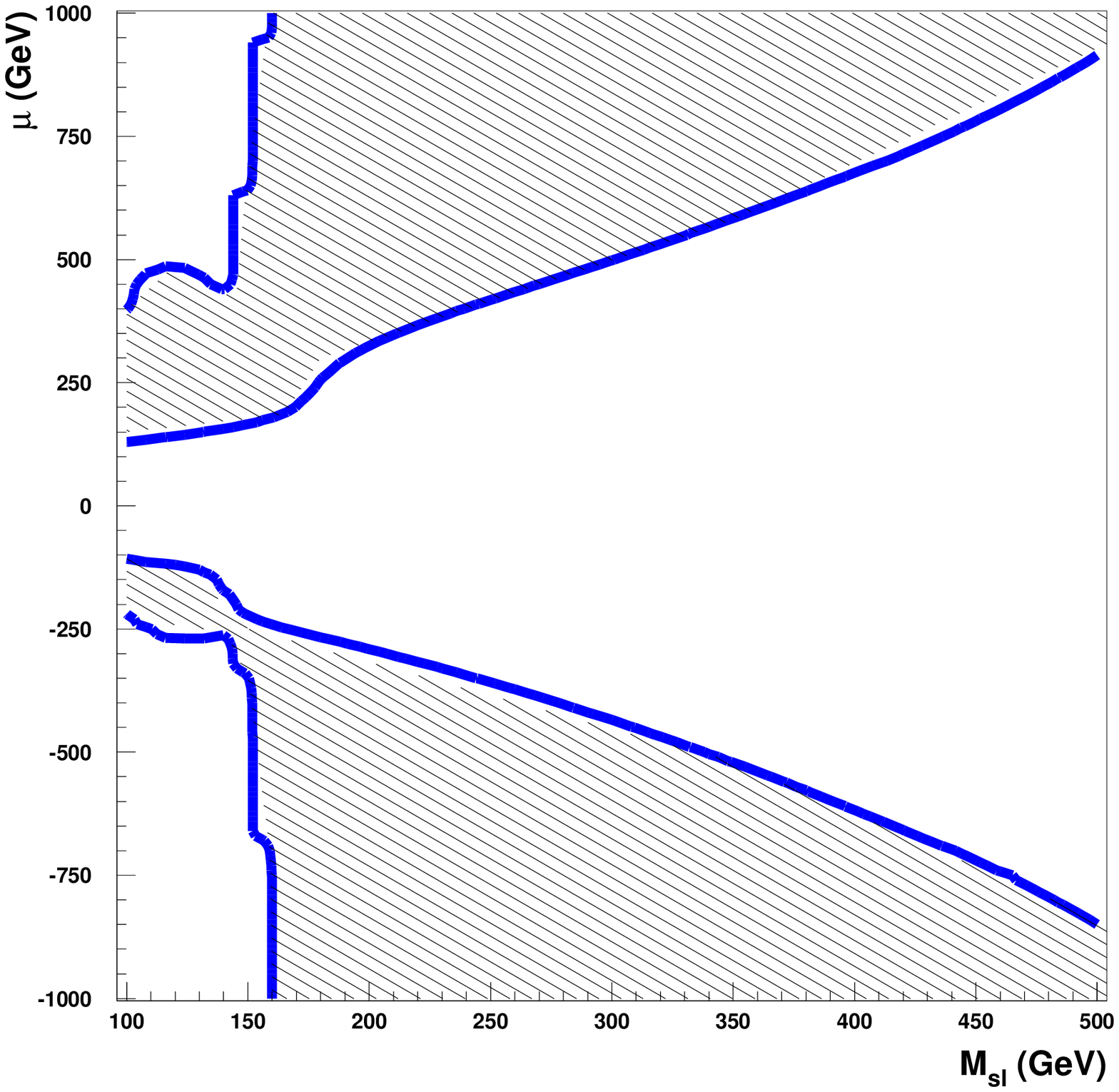}}
  \vspace{1.5cm}
  \caption{5$\sigma$-discovery contour in the $\mu-m_{\tilde{l}}$ plane for 100 $fb^{-1}$. \hspace{8cm}
  The other MSSM parameters are $\tan\beta$ = 5, $m_A$ = 350 GeV, $M_1$ = 80 GeV, $M_2$ = 150 GeV, 
  $m_{\tilde{g}, \tilde{q}}$ = 1000 GeV.}  
  \label{fig:dr2} 
  \end{center}
\end{figure}

The Higgsino mass parameter $\mu$ (together with the wino mass $M_2$) 
determines the gaugino/higgsino composition of the neutralinos in the 
intermediate state. For low values of $|\mu|$, the $\chi_2^0$ is rather 
higgsino-like and it will preferably decay into jets rather than into leptons.
For higher values of $|\mu|$, the gaugino content will dominate and they will
prefer to decay into leptons.
In fig. \ref{fig:dr2}, the 5$\sigma$-discovery contours in the $\mu$ - $m_{\tilde{l}}$ plane are 
shown for $\tan\beta$ = 5, $m_A$ = 350 GeV, $M_1$ = 80 GeV, $M_2$ = 150 GeV, 
$m_{\tilde{g}, \tilde{q}}$ = 1000 GeV. The indicated slepton mass is the right-handed slepton mass, 
the left-handed slepton mass is assumed to be $m_{\tilde{l}_R}$ + 50 GeV. 
\\ 
Within the discovery contours, one can distinguish two regimes. If $m_{\tilde{l}}$ 
$<$ 150 GeV ( $\approx$ $M_2$), decays of neutralinos in real sleptons are
allowed. This results in very high branching ratios in the range 200 $\lesssim$ $\mu$ $\lesssim$ 500 GeV
(-250 $\lesssim$ $\mu$ $\lesssim$ -100 GeV for negative values); for $\mu$ values outside this range or 
if the splitting between $m_{\tilde{l}_R}$ and $m_{\tilde{l}_L}$ is not sufficient, 
the neutralinos will preferably decay into sneutrino-neutrino pairs rather
than into slepton-lepton pairs, thus leading to pure $E_T^{miss}$.\\ 
For $m_{\tilde{l}}$ $>$ 150 GeV ( $\approx$ $M_2$), only virtual slepton exchange is
allowed. In this case, the branching ratio decreases steadily with the mass of
the sleptons. Negative values of $\mu$ seem slightly more favourable. 

\clearpage   
{\bf{The bino mass $M_1$ and wino mass $M_2$}}

If neutralinos are gaugino-like (large $|\mu|$ regime), the mass of the $\chi^0_1$ will 
approximately be equal to the bino mass parameter $M_1$
and the mass of the $\chi^0_2$ will be about the same as the wino mass parameter $M_2$.
Therefore, $M_2$ will determine the kinematic threshold from where the discovery reach
starts: $m_A \ge 2 M_2$. From fig. \ref{fig:dr1} it is clear that low values of $M_2$ and $M_1$ are much more favourable
in order to obtain a large discovery region in $m_A$. \\
The analysis presented in the previous paragraphs assumes that $M_2 - M_1$ $<$ $m_{Z^0}$. 
If this is not the case the neutralinos will decay mainly into real $Z^0$ bosons, 
leading to a 6 \% branching ratio into leptons. Furthermore the Z veto would reject these dilepton pairs. 
Therefore the upper limits on $M_1$ and $M_2$ for the present analysis to be valid 
are $M_1$ $\approx$ 90 GeV, $M_2$ $\approx$ 180 GeV (depending on $\mu$ and assuming the gaugino mass unification relation $M_2 \approx 2 M_1$). 
If $m_{\chi^0_2} - m_{\chi^0_1}$ $\ge$ $m_{Z^0}$ we will be forced to review the
selection strategy and omit the $Z$ veto leading to a 
larger Standard Model $ZZ$ background since it can now only be rejected using
the $E_T^{miss}$ requirement.\\
In the case of non-universal gaugino masses, it is  necessary that $M_1$ and
$M_2$ are not to close to each other ($M_1 \lnapprox~M_2$) to allow for
leptons with $P_T$ $>$ 10 GeV.

{\bf{Effects of light squarks and gluinos}}

If squarks and gluinos are light, they will be copiously produced at the LHC,
thus providing a large extra background due to the cascade decays of the squarks
and gluinos into next-to-lightest neutralinos. However, the neutralinos produced in these cascade decays
will be accompanied with a large number of jets and a large amount of 
$E_T^{miss}$. 
The signal, often produced in association with a $b\bar{b}$ pair, will contain not more than these two soft, rather forward jets 
(plus eventual initial/final state radiation jets).
This feature allows us to discriminate between the signal and the
squark/gluino background. Therefore, if this background would become important, it can be drastically reduced by
selecting events with not more than two jets with $E_T$ $\lesssim$ 50 GeV and putting an upper limit on
$E_T^{miss}$. Nevertheless, since the
production cross section rises steeply with decreasing squark/gluino mass, 
one can see from table 2 that for masses lower than 500 GeV, the squark/gluino background starts to become 
comparable to the sneutrino pair production background and the discovery contours will shrink accordingly.
In table 2, the squark/gluino pair production is estimated for
different values for $m_{\tilde{q},\tilde{g}}$ and compared with a 350 GeV Higgs signal, after applying the extra cuts. 
MSSM parameters are taken the same as in table 1.

\begin{table}[htb]\label{tabb:2}
    \caption{Number of surviving background and signal events after successive selection cuts (for 100 $fb^{-1}$).}
    \begin{center}
    \begin{tabular}{|l||c|c|c|c||c|} \hline 
        Process           & 800 GeV $\tilde{q}$ \& $\tilde{g}$ & 600 GeV $\tilde{q}$ \& $\tilde{g}$ & 500
        GeV $\tilde{q}$ \& $\tilde{g}$ & 400 GeV $\tilde{q}$ \& $\tilde{g}$ & 350 GeV $H^0$/$A^0$   \\ \hline
     cuts as in table 1   &    0    &   12  &    94     &    670   &   164 \\ 
     $E_T^{jet} < 50 GeV$   &    0    &   3  &    6     &    25   &   133 \\ 
     max. 2 jets           &    0    &   1  &     2     &     8   &   130 \\ 
     \hline     
      \end{tabular}
    \end{center}
\end{table}

\section{Conclusion}\label{chap:concl}
The sparticle decay modes of the heavy neutral SUSY Higgs bosons have been investigated. \\
The channel $A^0, H^0 \rightarrow  \chi^0_2 \, \chi^0_2 \rightarrow 4 \, l^{\pm}$ ($l$ = $e$, $\mu$) seems the most promising. 
In a rather large region of the MSSM parameter space, a clean signal can be observed by 
selecting events with 4 isolated leptons in the final state.
The main backgrounds are $ZZ$ and sparticle pair production (sneutrino, neutralino), but they can
be sufficiently suppressed using appropriate selection criteria.
Extra backgrounds due to light squark/gluino production can also be kept under control by applying additional cuts. \\
In the most common case where direct decays of neutralinos to sleptons are not allowed,
the $\chi^0_2 \, \chi^0_2$ channel seems to provide  
a detectable signal in the region between $m_A$ $\approx$ 230 and 450 GeV and for $\tan\beta$ $\lesssim$ 40 (at 100 $fb^{-1}$), 
in a scenario where $M_2$ $\approx$ 120 GeV, $\mu$ $\approx$ -500 GeV
and $m_{\tilde{l}}$ $\approx$ 250 GeV. \\
Since the branching ratio of the $A^0, H^0$ into four leptons is determined by the interplay between 
a number of MSSM parameters, the observability will also depend strongly on the values of 
$\mu$, $m_{\tilde{l}}$, $M_1$ and $M_2$. Large values of $|\mu|$ and low values of $m_{\tilde{l}}$ 
are favourable since they enhance the decay rate of the neutralinos into leptons. \\
\\
Motivated by the low $\tan\beta$ discovery potential of the $A^0, H^0 \rightarrow  \chi^0_2 \, 
\chi^0_2 \rightarrow 4 \, l$ channel, we also plan a similar study of the sparticle decay modes of the charged 
Higgs bosons $H^{\pm}$.

\section{Acknowledgements}
The authors would like to thank Abdel Djouadi for helpful discussions.

\end{document}